\documentclass[12pt,a4paper]{article}
\usepackage{a4wide}
\usepackage{amsmath}
\usepackage{amssymb}
\usepackage{epsfig}
\usepackage{subfigure}
\usepackage{exscale}
\usepackage{float}
\usepackage{bbm}
\usepackage[numbers,sort&compress]{natbib}
\usepackage{pst-plot, pstricks,pst-math}
\usepackage{fancybox,amssymb,color}
\usepackage{graphicx}
\usepackage{pstricks, color, graphicx, epsfig, psfrag}
\usepackage{amsfonts,amsmath,amssymb,slashed}
\usepackage{dsfont}
\usepackage{bbm,bm}
\usepackage{fancyhdr, a4wide}
\usepackage[english]{babel}
\usepackage{subfigure}

\makeatletter
\@addtoreset{equation}{section}
\makeatother

\title{Square-Lattice Antiferromagnet Subjected to a Magnetic Field Aligned with the Order Parameter}

\author{Christoph P.\ Hofmann$^a$ \\ \\
\normalsize{$^a$ Facultad de Ciencias, Universidad de Colima} \\
\vspace{0.3cm}
\normalsize{Bernal D\'iaz del Castillo 340, Colima C.P.\ 28045, Mexico} \\}

\begin{document}
\maketitle

\begin{abstract} \normalsize

The thermal properties of antiferromagnetic films -- in particular, the square-lattice antiferromagnet -- subjected to an external magnetic
field pointing into the direction of the staggered magnetization are explored. The effective field theory analysis of the free energy
density is carried out to two-loop order. While the emphasis is on finite temperature, we also discuss the behavior of the magnetization
and staggered magnetization at zero temperature. Our results imply that the staggered magnetization increases in presence of the magnetic
field -- reminiscent of magnetic catalysis. Most remarkably, if staggered and magnetic field strength are kept fixed, the magnetization
initially grows when temperature increases.

\end{abstract}

\maketitle

\section{Introduction}
\label{Intro}

Antiferromagnetic films at finite temperature have been explored in many theoretical studies. Here we point to those articles that also
discuss the effect of an external magnetic field: Refs.~\citep{HH69,Gho73,AUW77,Fis89,Tak89,FKLM92,AS93,Glu93,HN93,MG94,SSS94,San99,SS02,
HSR04,VCC05,SSKPWLB05,HSSK07,TZSS08,KSHK08,CZ09,LL09,FZSR09,HRO10,SST11,SST13,PR15}. While conventional condensed matter approaches rely on
microscopic models, phenomenological considerations or Monte Carlo simulations -- among others -- here we use magnon effective field theory
that allows for a {\it systematic} analysis of the low-temperature behavior of antiferromagnetic films. Recently, the thermodynamics of
antiferromagnetic films in magnetic and staggered fields has been analyzed up to two-loop order within effective Lagrangians in
Refs.~\citep{Hof17,Hof18}. There, the external magnetic field was oriented perpendicular to the staggered magnetization vector.

On the other hand, in the present study we investigate the situation where the magnetic field is aligned with the staggered magnetization.
It should be noted that all previous studies on antiferromagnetic films in magnetic fields -- to the best of our knowledge -- consider
noninteracting magnons and hence neglect the role of the spin-wave interaction in the thermodynamic properties.\footnote{With the exception
of Refs.~\citep{Hof17,Hof18}.} A full-fledged systematic analysis of antiferromagnetic films subjected to an external magnetic field that
is aligned with the order parameter, seems to be lacking. The present effective field theory investigation closes this apparent gap in the
quantum magnetism literature, by taking the effective evaluation of the partition function up to the two-loop level.

We focus on the behavior of the staggered magnetization and the magnetization in presence of staggered and magnetic fields at finite, but
also at zero temperature. While our effective investigation applies to any bipartite two-dimensional lattice, our numerical analysis
concentrates on the square-lattice antiferromagnet where all relevant low-energy effective couplings have been determined by Monte Carlo
simulations. As a consequence, the effective theory results are parameter-free and fully predictive. We observe that the staggered
magnetization grows when the staggered or the magnetic field become stronger -- the latter is reminiscent of magnetic catalysis. The
magnetization behaves in a similar way: it rises when magnetic and staggered field strength augment. But most remarkably, if magnetic and
staggered fields are kept constant, the magnetization {\it increases} when temperature is raised. The magnetization however starts to
decrease at more elevated temperatures -- as one would expect.

The paper is organized as follows. In Sec.~\ref{MicroEff} we provide a concise overview of the effective field theory description of
antiferromagnetic films. In particular, we derive the dispersion relations and construct the corresponding thermal propagators for magnons
in magnetic fields aligned with the staggered magnetization. Sec.~\ref{FreeEnergyDensity} is devoted to the evaluation of the free energy
density that we take up to the two-loop level. The behavior of the staggered magnetization and the magnetization in staggered and magnetic
fields is discussed in Sec.~\ref{LowTSeries}, both for zero and finite temperature. In Sec.~\ref{conclusions} we then conclude. A few
technical details concerning the perturbative evaluation of the partition function within effective field theory are presented in an
appendix.

\section{Microscopic and Effective Description}
\label{MicroEff}

On the microscopic level, the starting point to describe antiferromagnetic films is the quantum Heisenberg model extended by external
magnetic (${\vec H}$) and staggered (${\vec H_s}$) fields, 
\begin{equation}
\label{HeisenbergZeemanH}
{\cal H} \, = \, - J \, \sum_{n.n.} {\vec S}_m \! \cdot {\vec S}_n \, - \, \sum_n {\vec S}_n \cdot {\vec H} \, - \, \sum_n (-1)^n {\vec S}_n
\! \cdot {\vec H_s} \, , \qquad \qquad J < 0 \, , \quad J = const. 
\end{equation}
It is assumed that we are dealing with a bipartite lattice and that the sum in the first term is over nearest neighbor spins. If no
external fields are present, we have two magnon (Goldstone) modes -- or spin-wave branches -- that are degenerate and satisfy the
dispersion relation
\begin{equation}
\label{disprelAF}
\omega(\vec k) \, = \, v|{\vec k}| + {\cal O}({\vec k}^3) \, , \qquad {\vec k} = (k_1,k_2) \, ,
\end{equation}
where $v$ is the spin-wave velocity. If external fields are included, then the spontaneously broken O(3) symmetry of the isotropic
Heisenberg Hamiltonian is no longer exact, and the dispersion relations become gapped (see below).

On the effective field theory level\footnote{More detailed presentations of the effective description of antiferromagnetic materials in
magnetic and staggered fields have been given, e.g., in sections IX-XI of Ref.~\citep{Hof99a}. We also refer to the more conceptual
articles \citep{Leu94a,ABHV14} that deal with the foundations of effective Lagrangian field theory in condensed matter physics. In the
present article, we restrict ourselves to the most basic ingredients.}, the two antiferromagnetic magnon fields -- $U^1$ and $U^2$ -- are
collected in the unit vector $U^i$,
\begin{equation}
U^i = (U^0, U^a) \, , \quad U^0 = \sqrt{1 - U^a U^a} \, ,\qquad a = 1,2 \, , \quad i = 0,1,2 \, .
\end{equation}
The antiferromagnetic ground state is given by ${\vec U}_0 = (1,0,0)$, while the magnons are interpreted as fluctuations of the vector
$\vec U$ in the two directions orthogonal to ${\vec U}_0$.

The effective field theory captures the physics of the system at low energies and relies on an expansion in powers of momenta (energy,
temperature), which is reflected in the effective Lagrangian through a derivative expansion. The leading contribution -- ${\cal L}^2_{eff}$
-- exhibits two space-time derivatives (momentum order $p^2$),
\begin{equation}
\label{Leff2}
{\cal L}^2_{eff} = \mbox{$ \frac{1}{2}$} F^2 D_{\mu} U^i D^{\mu} U^i + M_s H^i_s U^i \, .
\end{equation}
The covariant time and space derivatives are
\begin{equation}
D_0 U^i = {\partial}_0 U^i + {\varepsilon}_{ijk} H^j U^k \, , \qquad D_r U^i = {\partial}_r U^i \, , \qquad (r=1,2) \, .
\end{equation}
The magnetic field $H^i$ shows up in the time covariant derivative $D_0 U^i$, while the staggered field $H^i_s$ comes with the low-energy
effective constant $M_s$: this is the staggered magnetization at zero temperature and infinite volume. The square of the other low-energy
effective constant $F$ is identified with the spin stiffness $\rho_s = F^2$ (see Ref.~\citep{HL90}).

The next-to-leading order effective Lagrangian (momentum order $p^4$) takes the form
\begin{eqnarray}
\label{Leff4}
{\cal L}^4_{eff} & = & e_1 (D_{\mu} U^i D^{\mu} U^i)^2 + e_2 (D_{\mu} U^i D^{\nu} U^i)^2
+ k_1 \frac{M_s}{\rho_s} (H_s^i U^i) (D_{\mu} U^k D^{\mu} U^k) \nonumber \\
& & + k_2 \frac{M_s^2}{\rho_s^2} (H_s^i U^i)^2 + k_3 \frac{M_s^2}{\rho_s^2} H_s^i H_s^i \, .
\end{eqnarray}
It involves a total of five next-to-leading order low-energy effective constants. For the effective field theory to be predictive, the
numerical values of $e_1, e_2, k_1, k_2, k_3$ have to be known -- or at least, their order of magnitude has to be estimated (see below).

In the present study we consider the case where the magnetic field is aligned with the staggered field,
\begin{equation}
\label{externalFields}
{\vec H}_{||} = (H,0,0) \, , \qquad {\vec H}_s = (H_s,0,0) \, , \qquad H, H_s > 0 \, .
\end{equation}
Note that the direction of the staggered and magnetic field coincides with the direction of the staggered magnetization vector
${\vec U}_0$. These external fields induce an energy gap in the magnon dispersion relations, as we now show.

The leading order effective Lagrangian ${\cal L}^2_{eff}$ -- Eq.~(\ref{Leff2}) -- gives rise to the following terms quadratic in the magnon
fields $U^a \, (a=1,2)$,
\begin{equation}
\mbox{$ \frac{1}{2}$} \rho_s \partial_{\mu} U^a \partial^{\mu} U^a - \mbox{$ \frac{1}{2}$} \rho_s M^2 U^a U^a
- \rho_s H \epsilon_{ab} \partial_0 U^a U^b + \mbox{$ \frac{1}{2}$} \rho_s H^2 U^a U^a \, ,
\end{equation}
where the "magnon mass" $M$ is associated with the staggered field through
\begin{equation}
M^2 = \frac{M_s H_s}{\rho_s} \, .
\end{equation}
Defining two new independent magnon fields $u(x)$ and $u^{*}(x)$ as
\begin{equation}
\label{physicalMagnons}
u = U^1 + i U^2 \, , \qquad u^{*} = U^1 - i U^2 \, ,
\end{equation}
we obtain the equations of motion,
\begin{eqnarray}
& & \Box u + M^2 u + 2 i H {\dot u} - H^2 u = 0 \, , \nonumber \\
& & \Box u^{*} + M^2 u^{*} - 2 i H {\dot u^{*}} - H^2 u^{*} = 0 \, .
\end{eqnarray}
Accordingly, the two magnons, subjected to a magnetic field pointing into the same direction as the staggered field, obey the dispersion
relations
\begin{eqnarray}
\label{disprelAFHparallel}
\omega_{+} & = & \sqrt{{\vec k \,}^2 + \frac{M_s H_s}{\rho_s}} + H \, , \nonumber \\
\omega_{-} & = & \sqrt{{\vec k \,}^2 + \frac{M_s H_s}{\rho_s}} - H \, .
\end{eqnarray}
This is perfectly consistent with the condensed matter literature (see, e.g., Refs.~\citep{ABK61a,Nol86}). In the absence of external
fields, the above dispersion relations reduce to the linear ungapped dispersion relation Eq.~(\ref{disprelAF}).\footnote{Notice that we
have put the spin-wave velocity $v$ to one.}

It should be noted that the lower spin-wave branch $\omega_{-}$ becomes negative, unless the condition
\begin{equation}
\label{stabilityCondition}
H_s > \frac{\rho_s}{M_s} \, H^2
\end{equation}
is satisfied. The present analysis is based on the assumption that the above stability criterion is indeed met. Otherwise, if the magnetic
field becomes too strong compared to the staggered field, the staggered magnetization vector rotates into a direction perpendicular to the
magnetic field. This situation of mutually perpendicular magnetic and staggered fields has been considered in Refs.~\citep{Hof17,Hof18}
within effective field theory. In particular, in that case only one of the magnons "senses" the magnetic field
\citep{ABK61a,Hof17},
\begin{eqnarray}
\label{disprelAFHperp}
\omega_{I} & = & \sqrt{{\vec k}^2 + \frac{M_s H_s}{\rho_s} + H^2} \, , \nonumber \\
\omega_{I\!I} & = & \sqrt{{\vec k}^2 + \frac{M_s H_s}{\rho_s}} \, ,
\end{eqnarray}
and the dispersion relations maintain their relativistic structure for both magnons, the "magnon masses" amounting to
\begin{equation}
\label{masses}
M^2_{I} = \frac{M_s H_s}{\rho_s} + H^2 \, , \qquad M^2_{I\!I} = \frac{M_s H_s}{\rho_s} \, .
\end{equation}
On the other hand, if the magnetic field is aligned with the staggered magnetization, the dispersion relations are not relativistic
according to Eq.~(\ref{disprelAFHparallel}).

We now turn to the thermal propagators and the kinematical functions related to them. For antiferromagnetic magnons that obey the
dispersion relations Eq.~(\ref{disprelAFHparallel}), the propagators at zero temperature -- and in Euclidean space -- take the
form\footnote{Note that we only regularize in the spatial dimension $d_s$. The space-time dimension we denote as $d$ where $d=d_s+1$.}
\begin{eqnarray}
\label{regpropHd2}
\Delta^{\pm} (x) & = & \int \frac{\mbox{d} p_4}{2 \pi} \int \frac{{\mbox{d}}^{d_s} p}{{(2 \pi)}^{d_s}} \,
\frac{e^{i({\vec p} \, {\vec x} - p_4 x_4)}}{p_4^2 + {\vec p \,}^2 + M^2 \pm 2 i H p_4 - H^2} \nonumber \\
& = & {\int}_{\!\!\!0}^{\infty} \mbox{d} \lambda \, \int \frac{\mbox{d} p_4}{2 \pi} \int \frac{{\mbox{d}}^{d_s} p}{{(2 \pi)}^{d_s}} \,
e^{i({\vec p} \, {\vec x} - p_4 x_4)} e^{-\lambda (p_4^2 + {\vec p \,}^2 + M^2 \pm 2 i H p_4 - H^2)} \, .
\end{eqnarray}
In the present two-loop calculation, as will become clear in the next section, we only need the values of the propagators at the origin
$x$=0. Integration over Euclidean energy and momentum then leads to
\begin{eqnarray}
\Delta^{\pm} (0) & = & \frac{1}{2 \sqrt{\pi}} {\int}_{\!\!\!0}^{\infty} \mbox{d} \lambda \, \lambda^{-\frac{1}{2}}
\int \frac{{\mbox{d}}^{d_s} p}{{(2 \pi)}^{d_s}} \,e^{-\lambda ({\vec p \,}^2 + M^2)} \nonumber \\
& = & \frac{M^{d_s-1}}{2^{d_s+1} \pi^{\frac{d_s}{2}+\frac{1}{2}}} \, \Gamma\Big(-\frac{d_s}{2}+\frac{1}{2}\Big) \, .
\end{eqnarray}
Remarkably, upon integrating over $p_4$, the dependence on the magnetic field drops out: the propagator ${\Delta}^{+}$ that describes magnon
$u$ is identical with the propagator ${\Delta}^{-}$ that describes magnon $u^{*}$, and they furthermore coincide with the
(pseudo-)Lorentz-invariant and degenerate propagator $\Delta$,
\begin{equation}
\label{regprop}
\Delta(0) = \int \frac{{\mbox{d}}^d p}{{(2 \pi)}^d} \, \frac{1}{M^2 + p^2}
= {\int}_{\!\!\!0}^{\infty} \mbox{d} \lambda \, (4 \pi \lambda)^{-d/2} e^{-\lambda M^2} \, .
\end{equation}
The physical limit $d_s \to 2$ ($d \to 3$) is unproblematic and yields
\begin{equation}
\lim_{d_s \to 2} \Delta^{\pm}(0) = - \frac{M}{4 \pi} \, .
\end{equation}

The thermal propagators imply infinite sums and are constructed from the zero-temperature propagators as\footnote{For a brief account on
finite-temperature effective field theory, see Sec.~III of Ref.~\citep{Hof17}. Details on finite-temperature field theory are provided in
the textbook by Kapusta and Gale, Ref.~\citep{KG06} (chapters 2 and 3).}
\begin{equation}
\label{ThermalPropagators}
G^{\pm}(x) = \sum_{n = - \infty}^{\infty} \Delta^{\pm}({\vec x}, x_4 + n \beta) \, , \qquad \beta = \frac{1}{T} \, .
\end{equation}
Regularizing in the spatial dimensions only, they read
\begin{equation}
\label{ThermalPropagators2}
G^{\pm}(x) = \sum_{n = - \infty}^{\infty} \, {\int}_{\!\!\!0}^{\infty} \mbox{d} \lambda \, \int \frac{\mbox{d} p_4}{2 \pi}
\int \frac{{\mbox{d}}^{d_s} p}{(2 \pi)^{d_s}} \, e^{-ip_4 (x_4 + n \beta) + i {\vec p} \, {\vec x}} e^{-\lambda(p_4^2+{\vec p \,}^2 + M^2 \pm 2 i H p_4 - H^2)}
\, .
\end{equation}
Unlike at $T$=0, integration over Euclidean energy does not eliminate the magnetic field,
\begin{equation}
\label{ThermalPropagators3}
G^{\pm}(x) = \frac{1}{2 \sqrt{\pi}} \, \sum_{n = - \infty}^{\infty} \, {\int}_{\!\!\!0}^{\infty} \mbox{d} \lambda \, \int
\frac{{\mbox{d}}^{d_s} p}{(2 \pi)^{d_s}} \lambda^{-\frac{1}{2}} e^{-\lambda({\vec p \,}^2 + M^2)} e^{i {\vec p} \, {\vec x}}
e^{-\frac{{(x_4 + n \beta)}^2}{4 \lambda}} e^{\mp H (x_4 +n \beta)} \, .
\end{equation}
At the origin $x$=0, and in terms of the dimensionless parameters $h$ and $\tilde m$,
\begin{equation}
\label{hm}
h = \frac{1}{2 \sqrt{\pi}} \frac{H}{T} \, , \qquad {\tilde m} = \frac{1}{2 \sqrt{\pi}} \frac{M}{T} =
\frac{1}{2 \sqrt{\pi}} \frac{\sqrt{M_s H_s}}{\sqrt{\rho_s} T} \, ,
\end{equation}
the thermal propagators take the form
\begin{equation}
\label{ThermalPropagators4}
G^{\pm}(0) = \frac{T^{d_s-1}}{4 \pi} \sum_{n = - \infty}^{\infty} \, {\int}_{\!\!\!0}^{\infty} \mbox{d} \lambda \,
\lambda^{-\frac{d_s}{2}-\frac{1}{2}} e^{-\lambda {{\tilde m}^2}} e^{-\frac{\pi n^2}{\lambda}} e^{\mp 2 \sqrt{\pi} h n} \, .
\end{equation}
The infinite sum can be performed analytically with the result
\begin{equation}
\label{ThermalPropagatorsJacobi}
G^{\pm}(0) = \frac{T^{d_s-1}}{4 \pi} \, {\int}_{\!\!\!0}^{\infty} \mbox{d} \lambda \, \lambda^{-\frac{d_s}{2}} e^{-\lambda {\tilde m}^2} \,
\theta_3\Big( \pm \sqrt{\pi} h \lambda, e^{- \pi \lambda}  \Big) e^{\lambda h^2} \, ,
\end{equation}
where $\theta_3(u,q)$ is the Jacobi theta function defined by
\begin{equation}
\label{Jacobi3}
\theta_3(u,q) = 1 + 2 \sum_{n=1}^{\infty} q^{n^2} \cos(2 n u) \, .
\end{equation}
It should be noted that the thermal propagators $G^{+}(0)$ and $G^{-}(0)$ are in fact identical: the summation in
Eq.~(\ref{ThermalPropagators4}) is symmetrical and the Jacobi theta function is even in the parameter $u=\sqrt{\pi} h \lambda$. We
therefore adapt our notation by using
\begin{equation}
{\hat G(0)} = G^{+}(0) = G^{-}(0) \, .
\end{equation}
In order to isolate the purely thermal piece in ${\hat G}(0)$, we subtract the $n$=0 (zero-temperature) contribution from the infinite sum,
\begin{equation}
\label{ThermalPropagatorsg1}
{\hat g}_1 \equiv {\hat G}(0) - \Delta(0) \, .
\end{equation}
This then leads to the kinematical Bose function ${\hat g}_1$,
\begin{equation}
\label{g1Bose}
{\hat g}_1 = \frac{T^{d_s-1}}{4 \pi} \, {\int}_{\!\!\!0}^{\infty} \mbox{d} \lambda \, \lambda^{-\frac{d_s}{2}} e^{-\lambda {\tilde m}^2}
\Bigg\{ \theta_3\Big( \sqrt{\pi} h \lambda, e^{- \pi \lambda} \Big) e^{\lambda h^2} - 1 \Bigg\} \, .
\end{equation}
The limit $d_s \to 2$ in the above representation is well-defined and the numerical evaluation of
\begin{equation}
{\hat g}_1 = \frac{T}{4 \pi} \, {\int}_{\!\!\!0}^{\infty} \mbox{d} \lambda \, \lambda^{-1} e^{-\lambda {\tilde m}^2}
\Bigg\{ \theta_3\Big( \sqrt{\pi} h \lambda, e^{- \pi \lambda} \Big) e^{\lambda h^2} - 1 \Bigg\} \qquad \qquad (d_s = 2)
\end{equation}
can be done straightforwardly. For later purposes, it is convenient to define the dimensionless kinematical function ${\hat h}_1$ as
\begin{equation}
\label{defh1}
{\hat h}_1 = \frac{{\hat g}_1}{T} \, . 
\end{equation}

\section{Free Energy Density}
\label{FreeEnergyDensity}

\begin{figure}
\begin{center}
\includegraphics[width=15cm]{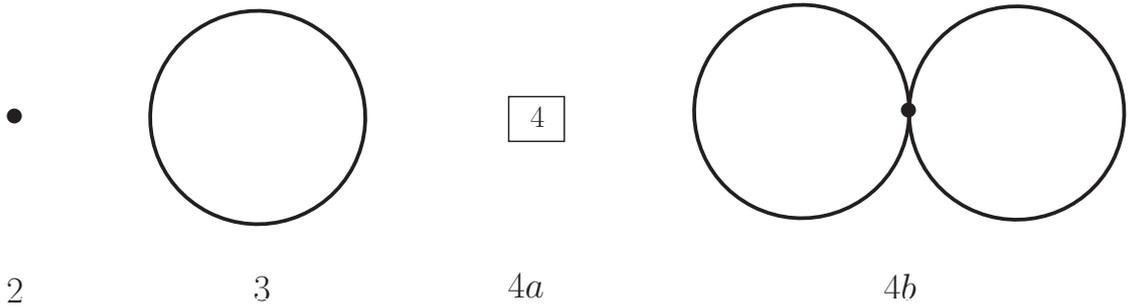}
\end{center}
\caption{Partition function diagrams up to order $T^4$ for antiferromagnets in two spatial dimensions. Filled circles stand for vertices
from ${\cal L}^2_{eff}$. The number 4 in the box stands for the subleading piece ${\cal L}^4_{eff}$.}
\label{figure1}
\end{figure}

We now evaluate the partition function -- or, equivalently, free energy density -- for antiferromagnetic films in presence of magnetic and
staggered fields aligned with the order parameter according to Eq.~(\ref{externalFields}). In Fig.~\ref{figure1} we depict the Feynman
diagrams that are relevant up to two-loop order.\footnote{More detailed information on the derivation of the partition function is given,
e.g., in section 2 of Ref.~\citep{Hof10} or in appendix A of Ref.~\citep{Hof11}.} The one-loop diagram $3$ describes the noninteracting
magnon gas and is of momentum (temperature) order $p^3$ ($T^3$). At next-to-leading order in the low-temperature expansion we have the
($T$-independent) tree graph $4a$ with an insertion from ${\cal L}^4_{eff}$, as well as the two-loop (interaction) graph $4b$ that involves
a vertex from the leading Lagrangian ${\cal L}^2_{eff}$. Both contributions are of order $p^4 \propto T^4$. Notice that the low-temperature
expansion is systematic: each loop referring to antiferromagnetic magnons is suppressed by one power of temperature in two spatial
dimensions (see Refs.~\citep{HN93,Hof10}).

The incorporation of a magnetic field aligned with the order parameter does not lead to additional vertices or Feynman diagrams: the set of
diagrams shown in Fig.~\ref{figure1} is the same also in the absence of ${\vec H_{||}}$. The parallel magnetic field only emerges
indirectly in the thermal propagators $G^{\pm}(x)$. In contrast, magnetic fields oriented {\it perpendicular} to the staggered magnetization
do generate new vertices with an odd number of magnon lines, yielding additional Feynman diagrams at two-loop order \citep{Hof17}.

The tree graphs $2$ and $4a$ that do not involve any magnon propagators, only give rise to zero-temperature contributions to the free
energy density,
\begin{eqnarray}
\label{z2z4a}
z_2 & = & - M_s H_s \, , \nonumber \\
z_{4a} & = & -(k_2 + k_3) \frac{M^2_s H^2_s}{\rho_s^2} \, .
\end{eqnarray}
Remarkably, the parallel magnetic field does not show up in these expressions. This is again different from a magnetic field orthogonal to
the staggered magnetization, where both $z_2$ and $z_{4a}$ receive additional terms due to the magnetic field (see Ref.~\citep{Hof17}).

Next we consider the one-loop graph $3$. Details on its evaluation can be found in Appendix \ref{appendixA1}. Here we just provide the
result,
\begin{equation}
\label{z3}
z_3  = - \frac{M^{3/2}_s H^{3/2}_s}{6 \pi \rho_s^{3/2}} - {\hat g}_0 \, .
\end{equation}
The finite-temperature piece is encapsulated in the kinematical function ${\hat g}_0$ that is related to the kinematical function
${\hat g}_1$ of the preceding section via
\begin{equation}
\label{derg0}
{\hat g}_1 = - \frac{\mbox{d} {\hat g}_0}{\mbox{d} M^2} \, .
\end{equation}
Accordingly, in two spatial dimensions, it takes the form
\begin{equation}
\label{g0Bose}
{\hat g}_0 = T^3 \, {\int}_{\!\!\!0}^{\infty} \mbox{d} \lambda \, \lambda^{-2} e^{-\lambda {\tilde m}^2}
\Bigg\{ \theta_3\Big( \sqrt{\pi} h \lambda, e^{- \pi \lambda}  \Big) e^{\lambda h^2} - 1 \Bigg\} \qquad \qquad (d_s = 2) \, .
\end{equation}
Note that the magnetic field and the staggered field enter through the parameters $h$ and ${\tilde m}$ that we have defined in
Eq.~(\ref{hm}). Again it is useful to introduce the dimensionless kinematical function ${\hat h}_0$ as
\begin{equation}
\label{defh0}
{\hat h}_0 = \frac{{\hat g}_0}{T^3} \, .
\end{equation}

Finally, the evaluation of the two-loop graph $4b$ yields\footnote{For details see Appendix \ref{appendixA2}.}
\begin{eqnarray}
z_{4b} & = & \frac{H}{\rho_s} \, {\hat g}_1 \, \frac{\partial {\hat g}_0}{\partial H}
- \frac{\sqrt{M_s H_s} H}{4 \pi \rho_s^{3/2}} \, \frac{\partial {\hat g}_0}{\partial H}
- \frac{H^2}{\rho_s}{( {\hat g}_1)}^2
+ \frac{\sqrt{M_s H_s} H^2}{2 \pi \rho_s^{3/2}} \, {\hat g}_1 \nonumber \\
& & - \frac{M_s H_s H^2}{16 \pi^2 \rho_s^2} \, .
\end{eqnarray}
If the magnetic field is not present, graph $4b$ does not contribute to the free energy density, as is known from earlier studies
\citep{Hof10}.

Collecting results, the two-loop representation for the free energy density reads
\begin{equation}
z = z_0 - {\hat g}_0 + \frac{H}{\rho_s} \, {\hat g}_1 \, \frac{\partial {\hat g}_0}{\partial H}
- \frac{\sqrt{M_s H_s} H}{4 \pi \rho_s^{3/2}} \, \frac{\partial {\hat g}_0}{\partial H}
- \frac{H^2}{\rho_s}{( {\hat g}_1)}^2
+ \frac{\sqrt{M_s H_s} H^2}{2 \pi \rho_s^{3/2}} \, {\hat g}_1\, ,
\end{equation}
where the zero-temperature contribution $z_0$ is
\begin{equation}
z_0 = - M_s H_s - \frac{M^{3/2}_s H^{3/2}_s}{6 \pi \rho_s^{3/2}} - (k_2 + k_3) \frac{M^2_s H^2_s}{\rho_s^2}
- \frac{M_s H_s  H^2}{16 \pi^2 \rho_s^2} \, .
\end{equation}

Inspecting Fig.~\ref{figure1}, one notices that next-to-leading order effective constants only matter in the tree graph $4a$. In the
present evaluation that extends up to two-loop order, these constants are thus only relevant at zero temperature. The finite temperature
properties of the system are completely fixed by the leading piece ${\cal L}^2_{eff}$ of the effective Lagrangian that is (pseudo-)Lorentz
invariant pursuant to Eq.~(\ref{Leff2}). Note that (pseudo-)Lorentz invariance is an accidental symmetry, i.e., a symmetry that is not
present in the microscopic Heisenberg model, but emerges on the effective field theory level at leading order. This implies that the
specific geometry of the bipartite lattice (square, honeycomb) does not matter in the effective description at the order we are
considering: the general structure of the low-temperature series is identical for any of these bipartite lattices.

What differs from lattice to lattice, however, are the concrete values of the effective low-energy constants $\rho_s$ and $M_s$ that appear
in ${\cal L}^2_{eff}$. To be specific, we quote the numerical values that have been obtained with high-precision loop-cluster
simulations (see Refs.~\citep{GHJNW09,JKNW08}) for the square lattice,
\begin{equation}
\label{squareLEC}
\rho_s = 0.1808(4) J \, , \quad M_s = 0.30743(1) / a^2 \, , \quad v = 1.6585(10) J a \, ,
\end{equation}
and the honeycomb lattice,
\begin{equation}
\label{honeyLEC}
\rho_s = 0.102(2) J \, , \quad {\tilde M_s} = 0.2688(3) \, , \quad v = 1.297(16) J a \, ,
\end{equation}
where
\begin{equation}
{\tilde M_s} = \frac{ 3 \sqrt{3}}{4} \, M_s \, a^2.
\end{equation}
All results refer to spin one-half. The low-energy constants and the spin-wave velocity $v$ are measured in units of the exchange integral
$J$ and the lattice constant $a$.

\section{Antiferromagnetic Films at Low Temperatures}
\label{LowTSeries}

For the discussion of the thermal properties of the system, it is convenient to introduce the three dimensionless parameters,
\begin{equation}
\label{definitionRatios}
m \equiv \frac{\sqrt{M_s H_s}}{2 \pi \rho_s^{3/2}} \, , \qquad
m_H \equiv \frac{H}{2 \pi \rho_s} \, \, , \qquad
t \equiv \frac{T}{2 \pi \rho_s} \, ,
\end{equation}
that describe the physics of the system at low energies. They measure the strength of the external fields $H_s$ and $H$, as well as
temperature, in units of the exchange integral $J$. This is because the denominator,
\begin{equation}
2 \pi \rho_s \approx J \, ,
\end{equation}
is of the order of $J$ that defines the relevant microscopic scale. The low-energy effective field theory operates in the sector where the
parameters $m, m_H, t$ are small. More concretely, for illustrative purposes, in the plots below we will consider the parameter space
defined by
\begin{equation}
\label{domain}
T, \, H, \, M (\propto \sqrt{H_s}) \ \lesssim 0.8 \ J \, .
\end{equation}

We should keep in mind that the (weak) staggered and magnetic fields cannot take arbitrary values: rather, the stability criterion,
Eq.~(\ref{stabilityCondition}), has to be satisfied. In order to stay away from this instability, in the subsequent plots for the free
energy density, staggered magnetization and magnetization, we restrict the parameter space by choosing
\begin{equation}
m > m_H + \delta \, , \qquad  \delta = 0.2 \, .
\end{equation}
This guarantees that we are in the safe region where our effective analysis applies.

In terms of $m, m_H$ and $t$, the dimensionless kinematical functions ${\hat h}_0$ and ${\hat h}_1$ defined in Eq.~(\ref{defh0}) and
Eq.~(\ref{defh1}) take the form
\begin{eqnarray}
{\hat h}_0 & = & {\int}_{\!\!\!0}^{\infty} \mbox{d} \lambda \, \lambda^{-5/2} e^{-\lambda m^2/4 \pi t^2}
\Bigg\{ \sqrt{\lambda} \, \theta_3\Big( \frac{m_H \lambda}{2 t}, e^{- \pi \lambda} \Big) e^{m_H^2 \lambda/4 \pi t^2} - 1 \Bigg\} \, ,
\nonumber \\
{\hat h}_1 & = & \frac{1}{4 \pi} \, {\int}_{\!\!\!0}^{\infty} \mbox{d} \lambda \, \lambda^{-3/2} e^{-\lambda m^2/4 \pi t^2}
\Bigg\{ \sqrt{\lambda} \, \theta_3\Big( \frac{m_H \lambda}{2 t}, e^{- \pi \lambda} \Big) e^{m_H^2 \lambda/4 \pi t^2} - 1 \Bigg\} \, .
\end{eqnarray}
Note that these representations refer to $d_s$=2.

Let us first point out that the temperature-dependent two-loop corrections are small with respect to the temperature-dependent one-loop
contribution. This we illustrate by considering the free energy density where the low-temperature expansion amounts to
\begin{eqnarray}
\label{freeEnergyDensity}
& & z = z_0 + {\hat z}_1 \, T^3 + {\hat z}_2 \, T^4 + {\cal O}(T^5) \, , \nonumber \\
& & \quad {\hat z}_1 = - {\hat h}_0 \, , \\
& & \quad {\hat z}_2 = \Bigg( 2 \pi m_H t \, {\hat h}_1 \, \frac{\partial {\hat h}_0}{\partial m_H}
- \frac{m m_H}{2} \, \frac{\partial {\hat h}_0}{\partial m_H}
- \frac{2 \pi m^2_H}{t} \, {({\hat h}_1 )}^2
+ \frac{m m^2_H}{t^2} \, {\hat h}_1 \Bigg) \frac{1}{2 \pi \rho_s t} \, . \nonumber
\end{eqnarray}
The leading contribution (one-loop diagram $3$) is of order $T^3$, the next-to-leading contribution (two-loop diagram $4b$) is of order
$T^4$.

\begin{figure}
\begin{center}
\hbox{
\includegraphics[width=8.0cm]{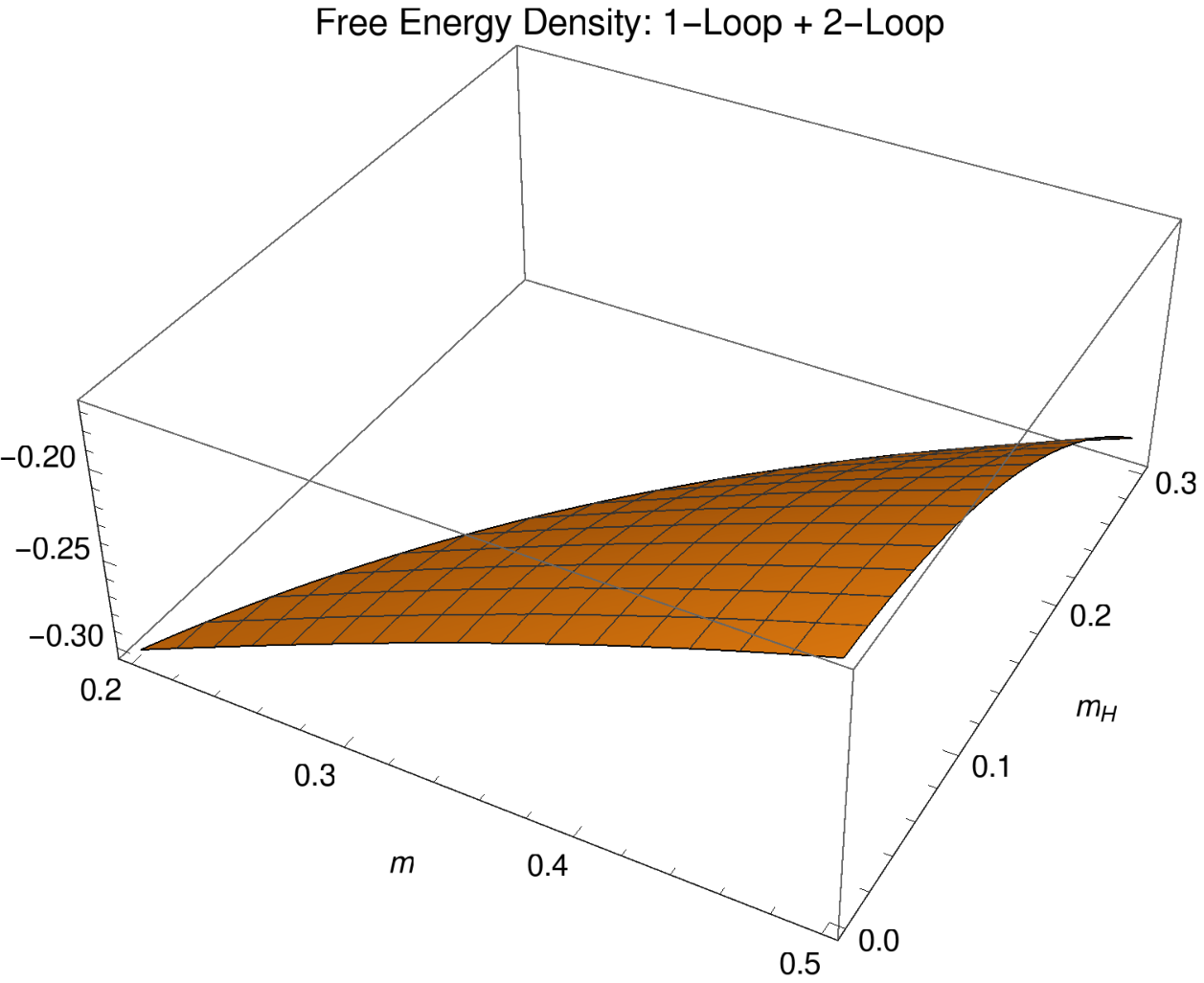} 
\includegraphics[width=8.0cm]{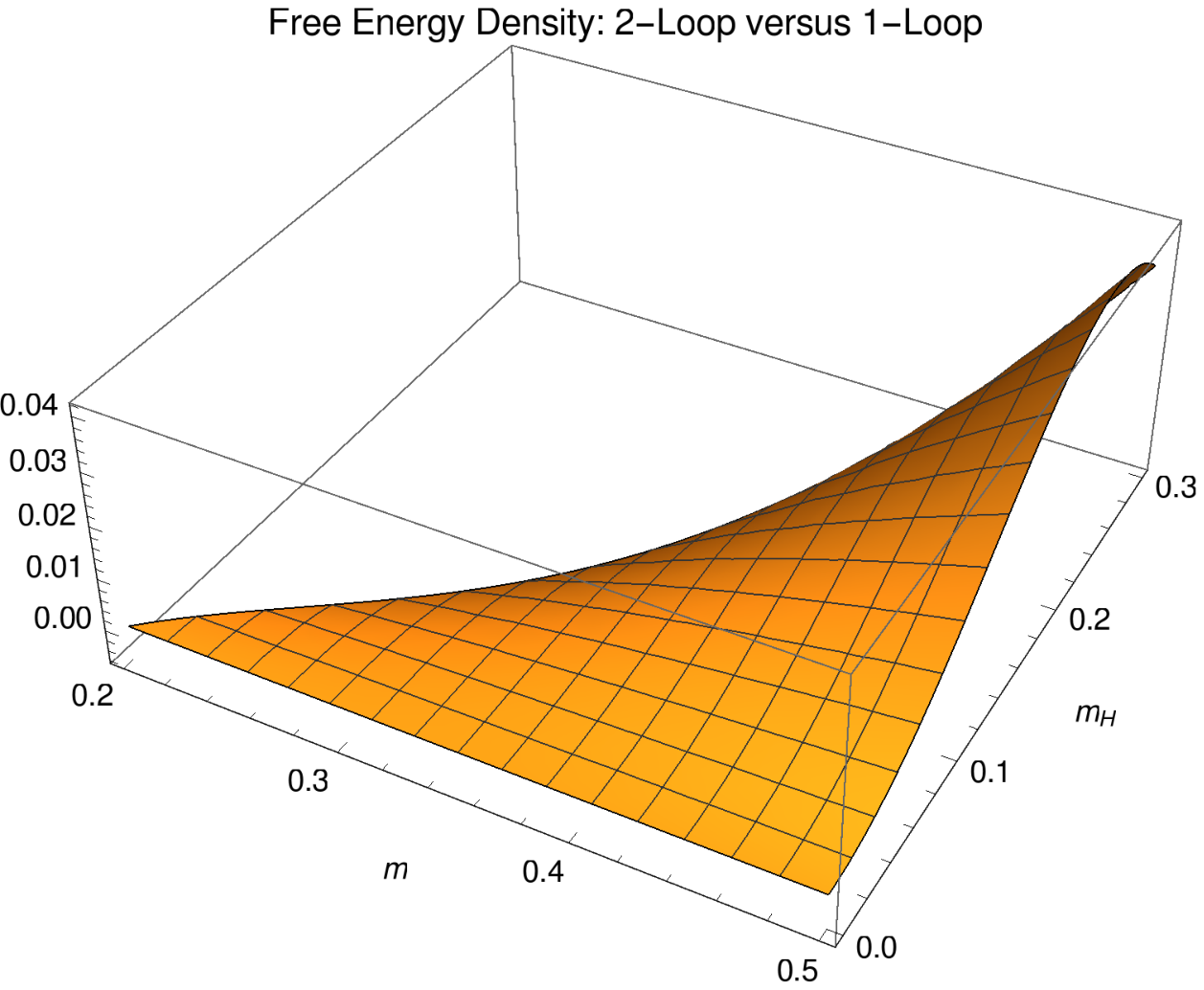}}
\vspace{7mm}
\hbox{
\includegraphics[width=8.0cm]{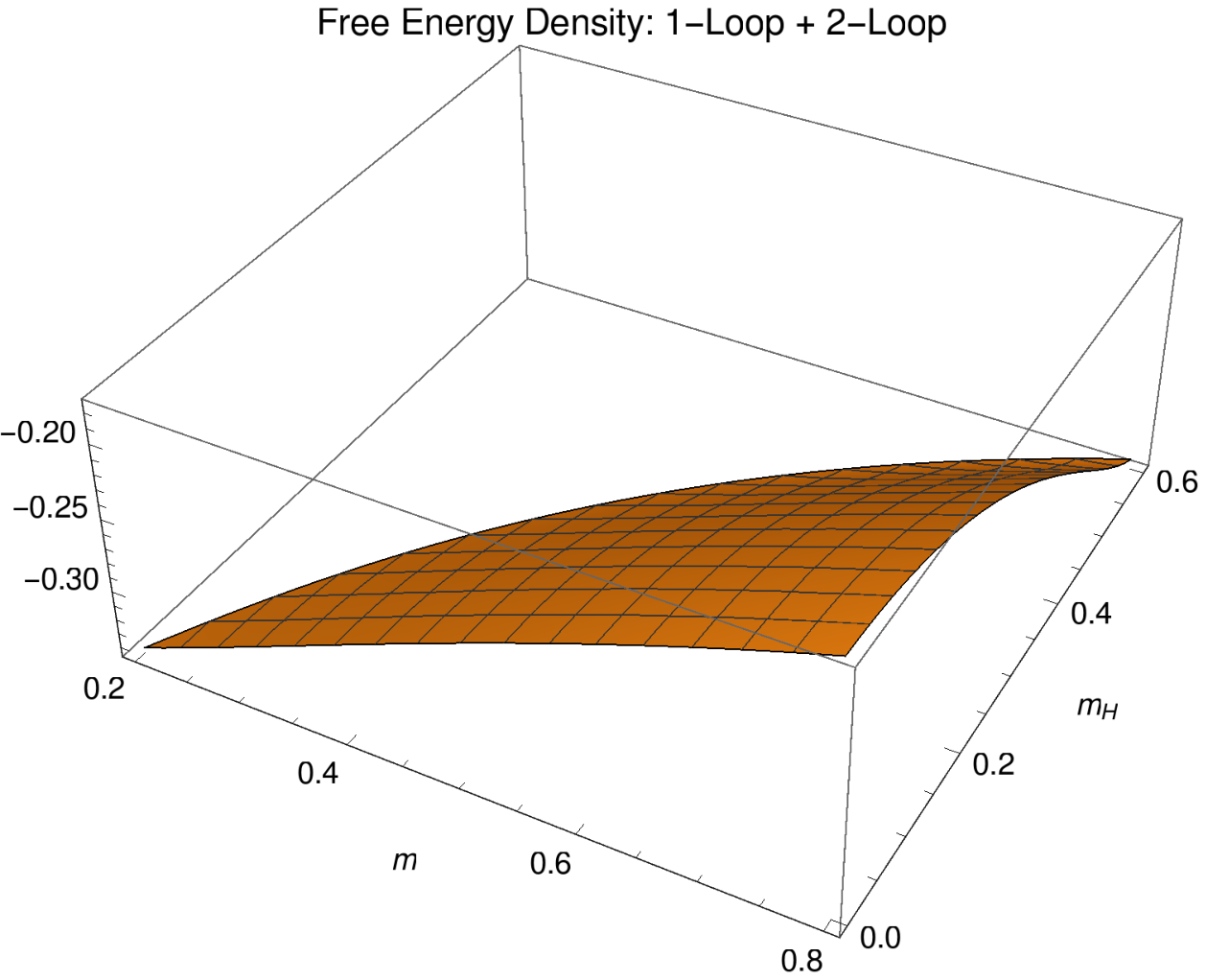}
\includegraphics[width=8.0cm]{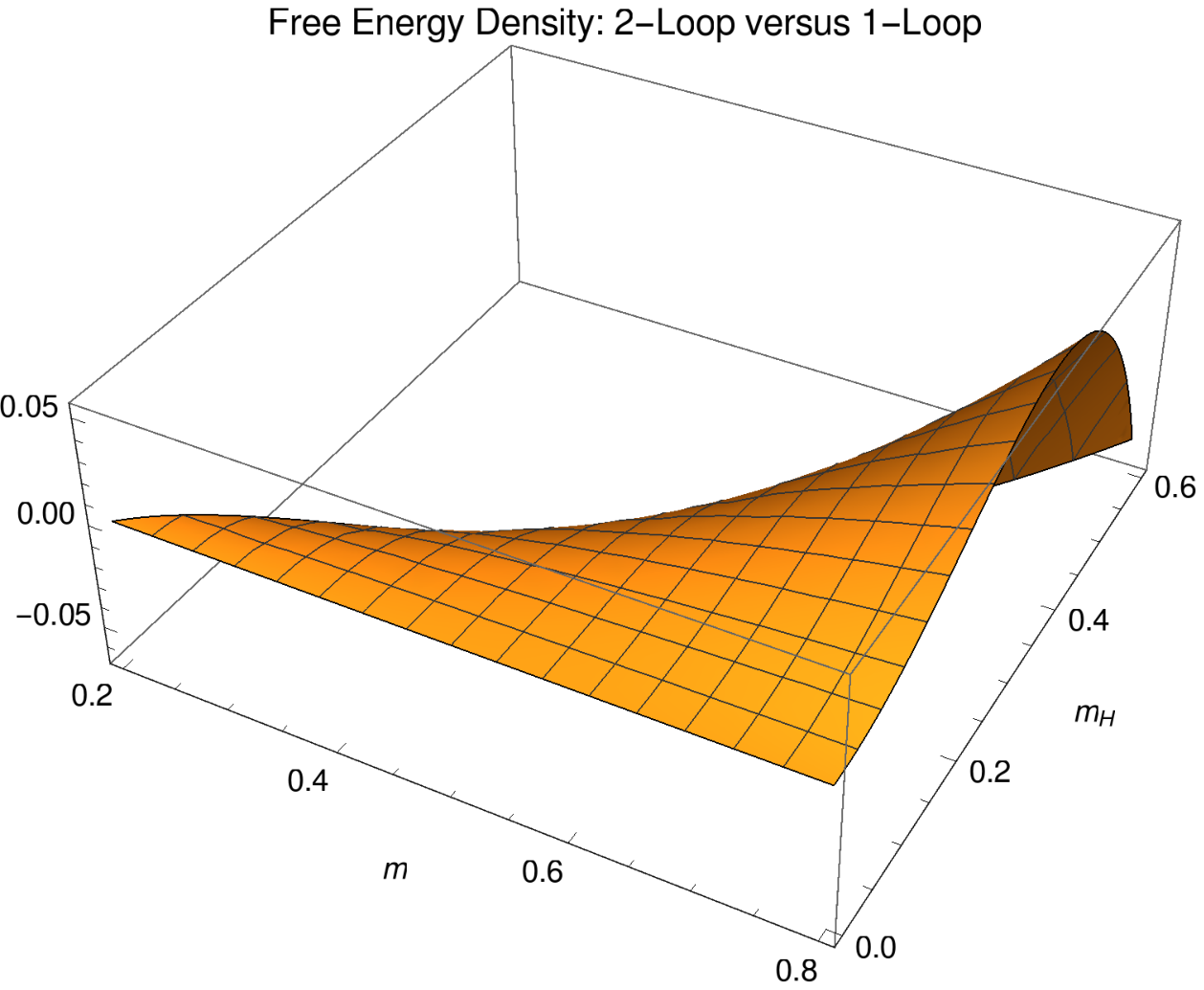}}
\end{center}
\caption{[Color online] Total temperature-dependent free energy density (LHS), Eq.~(\ref{figFED}), and two-loop versus one-loop
contribution (RHS), Eq.~(\ref{figFED2}), in magnetic and staggered fields at the temperatures $T/2 \pi \rho_s = 0.3$ (upper panel) and
$T/2 \pi \rho_s = 0.5$ (lower panel).}
\label{figure2}
\end{figure}

In Fig.~\ref{figure2}, on the respective left-hand sides, we depict the dimensionless quantity
\begin{equation}
\label{figFED}
\frac{{\hat z}_1 T^3 + {\hat z}_2 T^4}{T^3} \, ,
\end{equation}
i.e., the total temperature-dependent free energy density as a function of magnetic and staggered field strength for the two temperatures
$T/2 \pi \rho_s = 0.3$ (upper panel) and $T/2 \pi \rho_s = 0.5$ (lower panel). On the right-hand-sides of Fig.~\ref{figure2}, we then 
demonstrate that the one-loop contribution dominates the low-temperature expansion by plotting the dimensionless ratio
\begin{equation}
\label{figFED2}
\frac{{\hat z}_2 T}{{\hat z}_1}
\end{equation}
for the same two temperatures. Indeed, the two-loop corrections are small in either case. This is a generic result that also applies to the
magnetization and staggered magnetization we consider below. It shows that the effective low-temperature expansion is consistent: loops are
suppressed.

\subsection{Staggered Magnetization}
\label{stagMag}

The staggered magnetization can be extracted from the free energy density by
\begin{equation}
M_s(T,H_s,H) = - \frac{\partial z(T,H_s,H)}{\partial H_s} \, .
\end{equation}
The low-temperature series takes the structure
\begin{equation}
\label{OPAF}
M_s(T,H_s,H) = M_s(0,H_s,H) + {\tilde \sigma}_1 T + {\tilde \sigma}_2 T^2 + {\cal O}(T^3) \, ,
\end{equation}
where the coefficients are given by
\begin{eqnarray}
{\tilde \sigma}_1(T,H_s,H) & = & - \frac{M_s}{\rho_s} \, {\hat h}_1 \, , \nonumber \\
{\tilde \sigma}_2(T,H_s,H) & = & \frac{M_s}{\rho_s} \, \Bigg\{ \frac{m_H}{\rho_s} \, {\hat h}_2 \, \frac{\partial{\hat h}_0}{\partial m_H}
+ \frac{m_H}{\rho_s} \, {\hat h}_1 \, \frac{\partial{\hat h}_1}{\partial m_H}
+ \frac{m_H t}{8 \pi \rho_s m} \, \frac{\partial{\hat h}_0}{\partial m_H}  \nonumber \\
& & - \frac{m m_H}{4\pi \rho_s t} \, \frac{\partial{\hat h}_1}{\partial m_H}
- \frac{2 m_H^2}{\rho_s t^2} \, {\hat h}_1 {\hat h}_2
- \frac{m_H^2}{4\pi \rho_s m t} \, {\hat h}_1
+ \frac{m m_H^2}{2 \pi \rho_s t^3} \, {\hat h}_2 \Bigg\} \, .
\end{eqnarray}
The spin-wave interaction comes into play at order $T^2$. Again, in zero magnetic field, there is no interaction term at two-loop order:
${\tilde \sigma}_2(T,H_s,0) = 0$. The dimensionless kinematical function ${\hat h}_2$,
\begin{equation}
{\hat h}_2 = \frac{1}{16 \pi^2} \, {\int}_{\!\!\!0}^{\infty} \mbox{d} \lambda \, \lambda^{-1/2} e^{-\lambda m^2/4 \pi t^2}
\Bigg\{ \sqrt{\lambda} \, \theta_3\Big( \frac{m_H \lambda}{2t}, e^{- \pi \lambda} \Big)e^{m_H^2 \lambda/4 \pi t^2} - 1 \Bigg\} \, ,
\end{equation}
is defined as
\begin{equation}
{\hat h}_2 = {\hat g}_2 T \, ,
\end{equation}
where ${\hat g}_2$ can be obtained from ${\hat g}_1$ via
\begin{equation}
{\hat g}_2 = - \frac{\mbox{d} {\hat g}_1}{\mbox{d} M^2} \, .
\end{equation}
As we illustrate below, the behavior of the staggered magnetization in magnetic and staggered fields is dominated by the zero-temperature
contribution $M_s(0,H_s,H)$, i.e., the order parameter
\begin{equation}
M_s(0,H_s,H) = - \frac{\partial z(0,H_s,H)}{\partial H_s} = - \frac{\partial z_0}{\partial H_s}
\end{equation}
that amounts to
\begin{equation}
\label{OPT0}
\frac{M_s(0,H_s,H)}{M_s} = 1 + \frac{\sqrt{H_s M_s}}{4 \pi \rho_s^{3/2}} + \frac{H^2}{16 \pi^2 \rho_s^2}
+ \frac{2 (k_2 + k_3) H_s M_s}{\rho_s^2} \, .
\end{equation}
Here $M_s = M_s(0,0,0)$ is the order parameter with no external fields present. Note that, due to the tree graph $4a$, the combination
$k_2 + k_3$ of next-to-leading order effective coupling constants shows up in this temperature-independent piece. In the case of
antiferromagnetic films defined on a square lattice, this combination has been determined by very precise loop-cluster algorithms in
Ref.~\citep{GHJNW09} with the result\footnote{Notice that in Ref.~\citep{GHJNW09} a different convention for the low-energy constants was
used.}
\begin{equation}
\frac{k_2 + k_3}{v^2} = \frac{-0.0037}{2 \rho_s} = \frac{-0.0102}{J} \, .
\end{equation}
If the magnetic field is switched off, the $T$=0 staggered magnetization is governed by powers of $\sqrt{H_s}$, 
\begin{equation}
\label{OPinHszeroH}
M_s(0,H_s,0) = M_s + \frac{M_s^{3/2}}{4 \pi \rho_s^{3/2}} \, \sqrt{H_s } + \frac{2 M_s^2}{\rho_s^2} \, (k_2 + k_3) \, H_s
+ {\cal O}(H_s^{3/2}) \, .
\end{equation}
Remember that the other limit $H_s \to 0$ is not legitimate because it violates the stability condition (\ref{stabilityCondition}).

\begin{figure}
\begin{center}
\hbox{
\includegraphics[width=8.0cm]{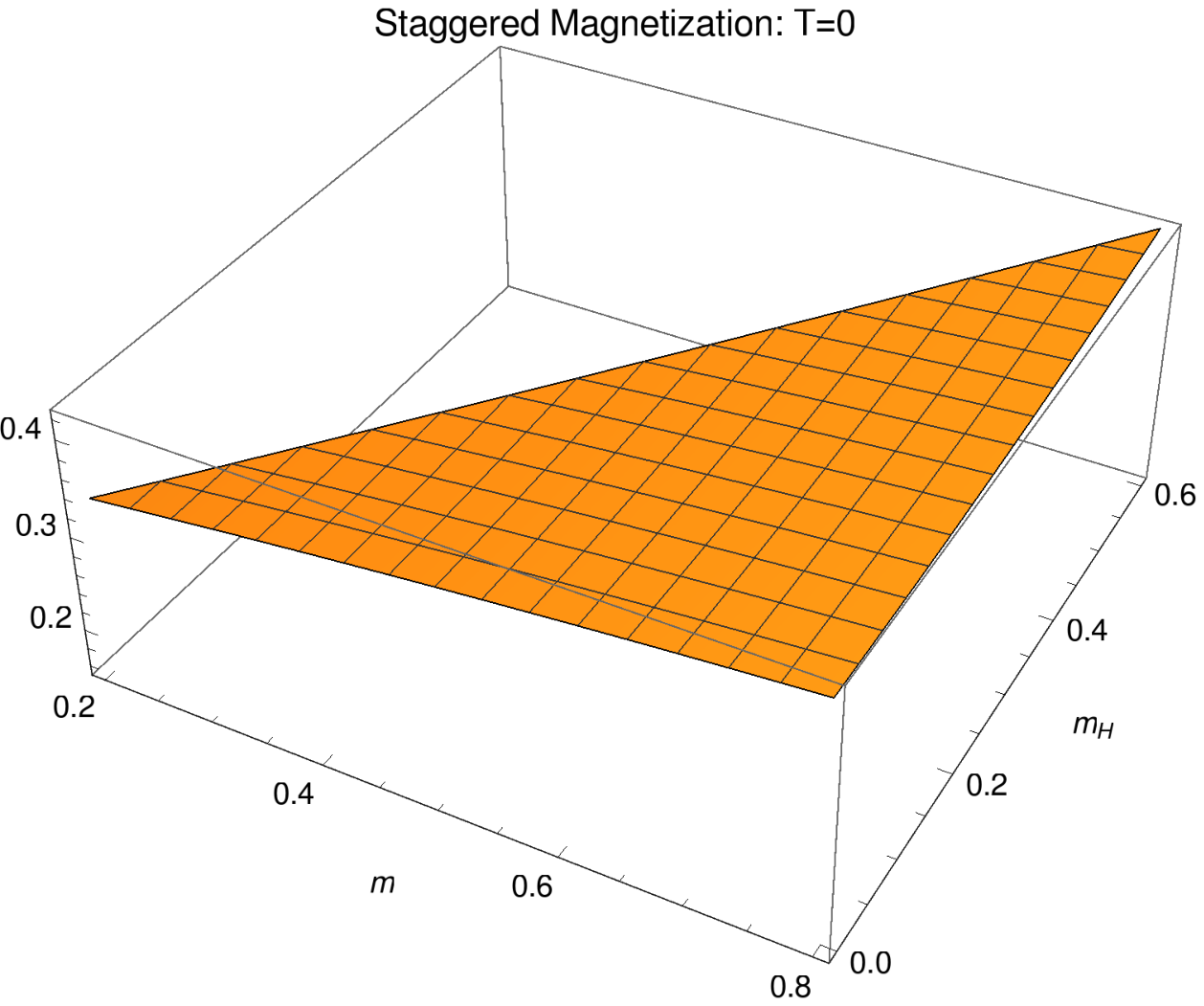} 
\includegraphics[width=8.0cm]{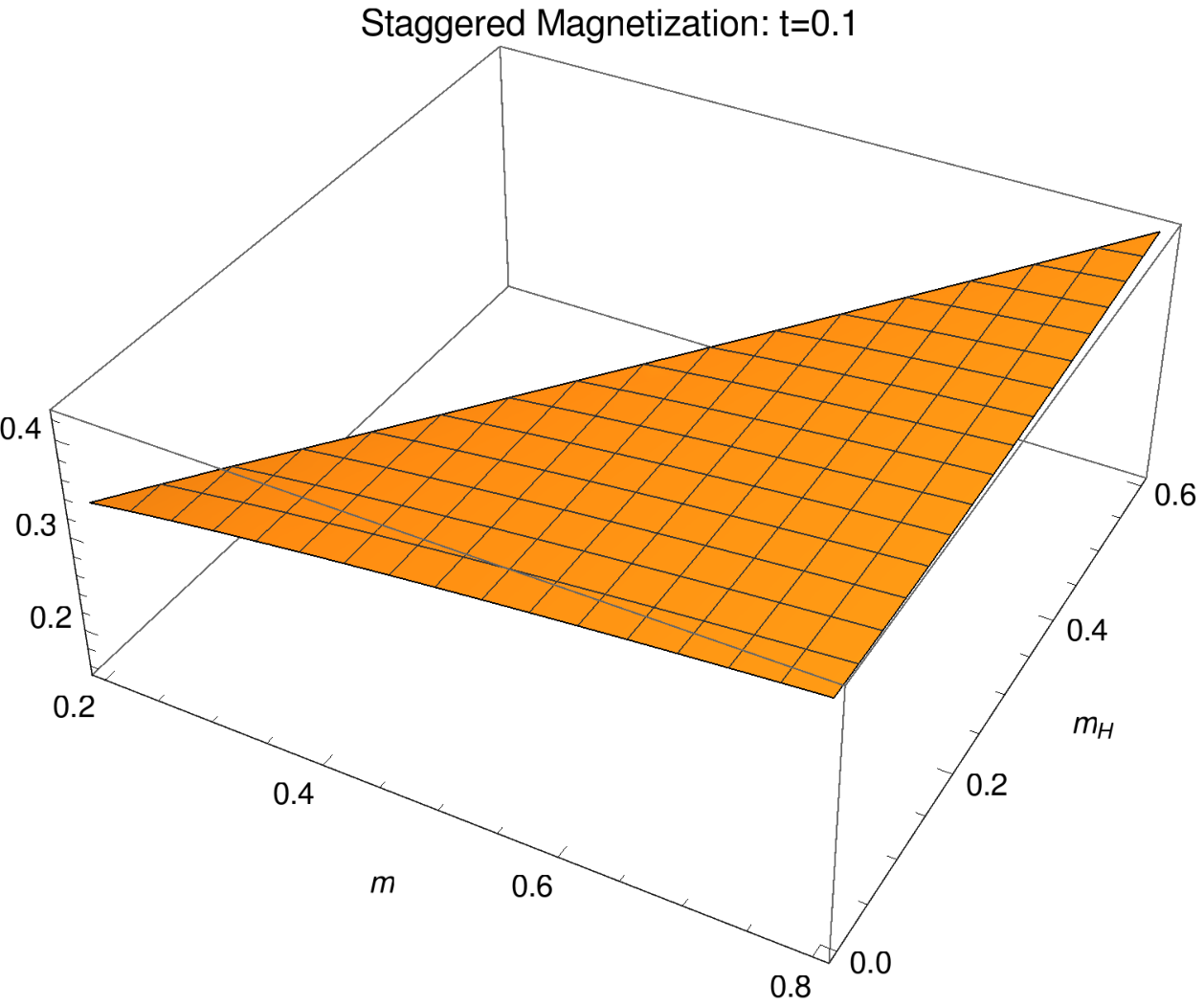}}
\vspace{7mm}
\hbox{
\includegraphics[width=8.0cm]{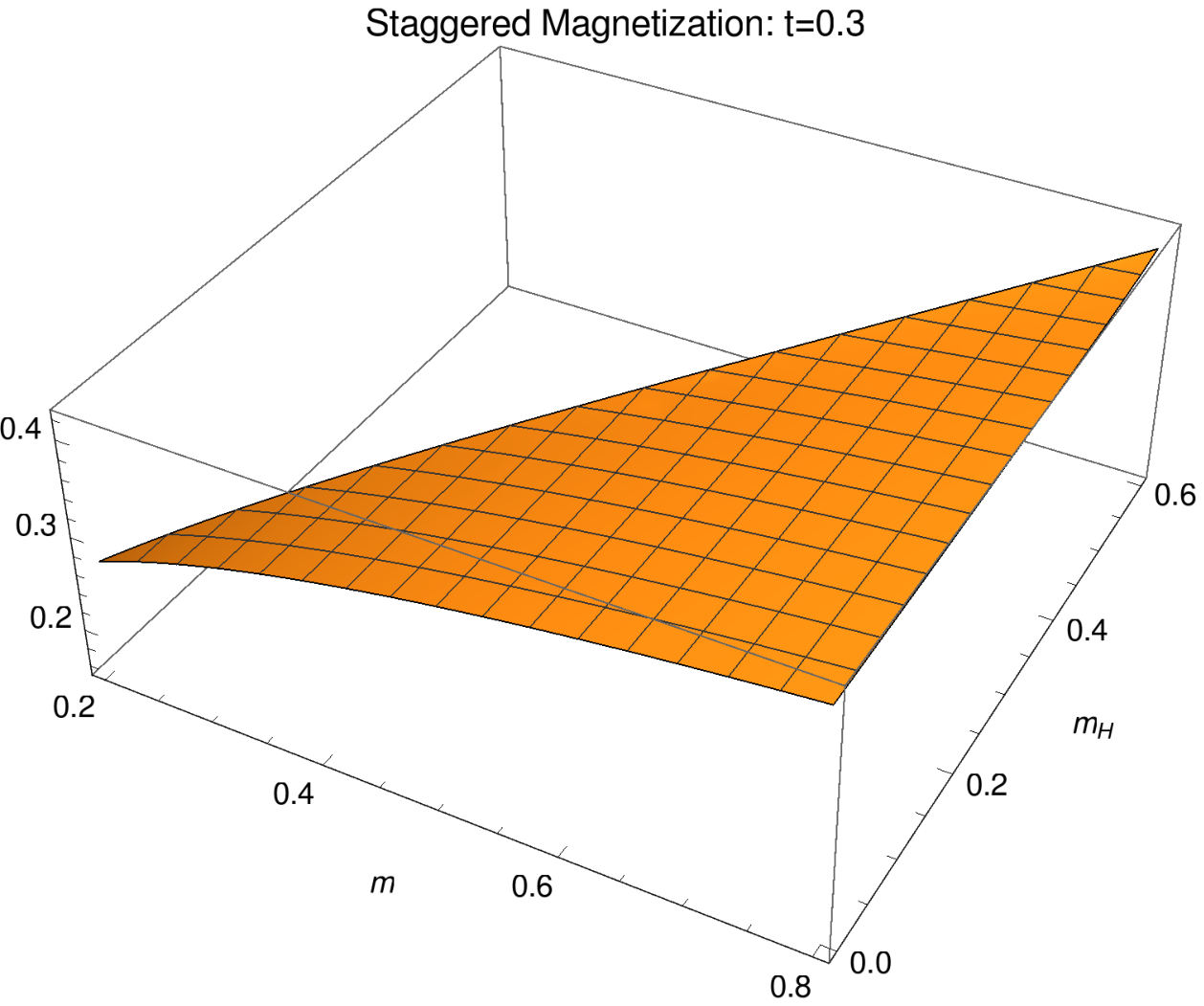}
\includegraphics[width=8.0cm]{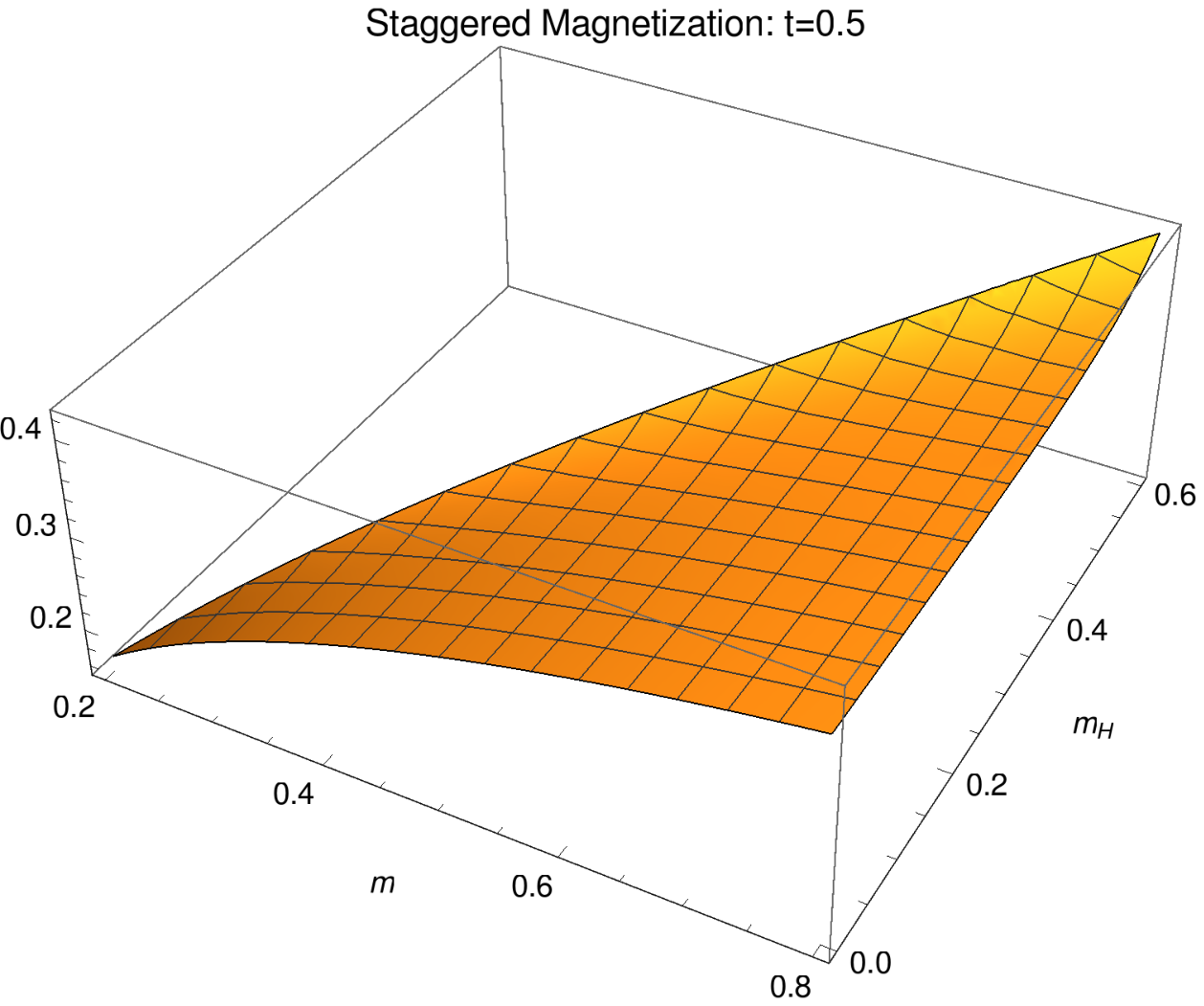}}
\end{center}
\caption{[Color online] Staggered magnetization $M_s(T,H_s,H)$ at zero and finite temperature as a function of magnetic ($m_H$) and
staggered ($m$) field strength for the square-lattice antiferromagnet. The upper left figure refers to $T$=0, the other figures refer to
the temperatures $t = T/2 \pi \rho_s = \{ 0.1, 0.3, 0.5 \}$ (left to right, top to bottom).}
\label{figure3}
\end{figure}

In Fig.~\ref{figure3} we examine the behavior of the staggered magnetization $M_s(T,H_s,H)$, Eq.~(\ref{OPAF}), in magnetic and staggered
fields, specifically for the square-lattice antiferromagnet where all relevant low-energy effective couplings are known. We first discuss
the zero-temperature case, i.e., the order parameter $M_s(0,H_s,H)$ that is depicted on the upper left. One notices that the order parameter
increases when the staggered field grows as one would expect: in stronger staggered fields the anti-alignment of the spins is more
pronounced. Remarkably, the order parameter also increases when the magnetic field gets stronger. This is reminiscent of magnetic catalysis
as described, e.g., in Refs.~\citep{ANT16,SS97,MS02,Oza14,Sho13,GMSS16}, and has also been observed in antiferromagnetic films where the
magnetic field is oriented {\it perpendicular} to the staggered magnetization (see Refs.~\citep{LL09,Hof17}). It should be noted that we
are dealing with an universal result -- not restricted to the square-lattice antiferromagnet -- because the coefficient involving the
magnetic field in Eq.~(\ref{OPT0}) is positive: irrespective of the actual value of the spin stiffness that indeed depends on the specific
bipartite lattice, the order parameter is enhanced when the magnetic field gets stronger.

The enhancement of the order parameter by magnetic and staggered fields can be explained by suppression of quantum fluctuations. The
staggered field, by construction, acts symmetrically on the two sublattices: it suppresses fluctuations of up-spins and down-spins in the
same manner, such that the staggered magnetization grows -- but the magnetization remains zero -- when only the staggered field is present.
If we now switch on a magnetic field pointing into the same direction as ${\vec H_s}$ on sublattice $A$, but pointing into the opposite
direction as ${\vec H_s}$ on sublattice $B$, the effect induced by the magnetic field is asymmetric: the net external field ($H_s+H$)
pointing up on sublattice $A$, is stronger than the net external field ($H_s-H$) pointing down on sublattice $B$. This leads to a
positive magnetization in the direction of the magnetic field (see next subsection) and, at the same time, it causes the order parameter to
rise.

Note that the suppression of quantum fluctuations is quite significant: in the absence of external fields, the staggered magnetization
of the square-lattice antiferromagnet takes the value $M_s = 0.30743(1) / a^2$. According to Fig.~\ref{figure3}, in the parameter region
we display ($m \le 0.8, m_H \le 0.6$), this value may increase up to $M_s \approx 0.42 / a^2$ in presence of the external fields.
Considering the fact that in the hypothetical configuration, where all spins would be perfectly antialigned, we would have
$M_s = \frac{1}{2} / a^2$, the effect we observe is quite large.

Let us now discuss the behavior of the staggered magnetization at finite temperature. Along with the $T$=0 contribution, in
Fig.~\ref{figure3}, we depict the staggered magnetization $M_s(T,H_s,H)$, Eq.~(\ref{OPAF}), for the three temperatures
$T/2 \pi \rho_s = \{ 0.1, 0.3, 0.5 \}$ (left to right, top to bottom). At low temperatures, the $T$=0 result is hardly modified. At more
elevated temperatures, the staggered magnetization may be reduced substantially, above all in weak magnetic and staggered fields where
thermal fluctuations win over the suppression of quantum fluctuations by the external fields. Overall, this is what one would expect
intuitively.

\subsection{Magnetization}

The low-temperature expansion of the magnetization,
\begin{equation}
M(T,H_s,H) = - \frac{\partial z(T,H_s,H)}{\partial H} \, ,
\end{equation}
takes the form
\begin{equation}
\label{magnetizationAF}
M(T,H_s,H) = M(0,H_s,H) + {\hat \sigma}_1 T + {\hat \sigma}_2 T^2 + {\cal O}(T^3) \, ,
\end{equation}
with coefficients
\begin{eqnarray}
{\hat \sigma}_1(T,H_s,H) & = & 2 \pi \rho_s t^2 \frac{\partial {\hat h}_0}{\partial m_H} \, , \nonumber \\
{\hat \sigma}_2(T,H_s,H) & = & - 2 \pi t^2 {\hat h}_1 \frac{\partial {\hat h}_0}{\partial m_H}
- 2 \pi m_H t^2 \frac{\partial {\hat h}_1}{\partial m_H} \frac{\partial {\hat h}_0}{\partial m_H}
- 2 \pi m_H t^2 {\hat h}_1 \frac{\partial^2 {\hat h}^2_0}{\partial m^2_H} \nonumber \\
& & + \frac{m t}{2} \frac{\partial {\hat h}_0}{\partial m_H}
+ \frac{m m_H t}{2} \frac{\partial {\hat h}^2_0}{\partial m^2_H}
+ 4 \pi m_H {({\hat h}_1)}^2
+ 4 \pi m^2_H {\hat h}_1 \frac{\partial {\hat h}_1}{\partial m_H} \nonumber \\
& & - \frac{2 m m_H}{t} {\hat h}_1
- \frac{m m_H^2}{t} \frac{\partial {\hat h}_1}{\partial m_H} \, .
\end{eqnarray}
The free Bose gas contribution is of order $T$, while the spin-wave interaction is contained in the $T^2$-term. The magnetization at zero
temperature is given by
\begin{equation}
\label{MagT0}
M(0,H_s,H) = \frac{H_s M_s H}{8 \pi^2 \rho_s^2} \, .
\end{equation}
In contrast to the order parameter, in the magnetization next-to-leading order effective constants are irrelevant: they only show up beyond
two loops. This means that our effective result for the total magnetization $M(T,H_s,H)$, Eq.~(\ref{magnetizationAF}), is fully predictive
also for the honeycomb-lattice antiferromagnet if the numerical values for the leading low-energy effective constants provided in
Eq.~(\ref{honeyLEC}) are inserted. Note that the limit $H_s \to 0$ in Eq.~(\ref{MagT0}) is not legitimate as it would violate the stability
criterion. On the other hand, if the magnetic field is switched off, the magnetization drops to zero
\begin{equation}
\lim_{H \to 0} M(0,H_s,H) = 0 \, ,
\end{equation}
as it should.

\begin{figure}
\begin{center}
\hbox{
\includegraphics[width=8.0cm]{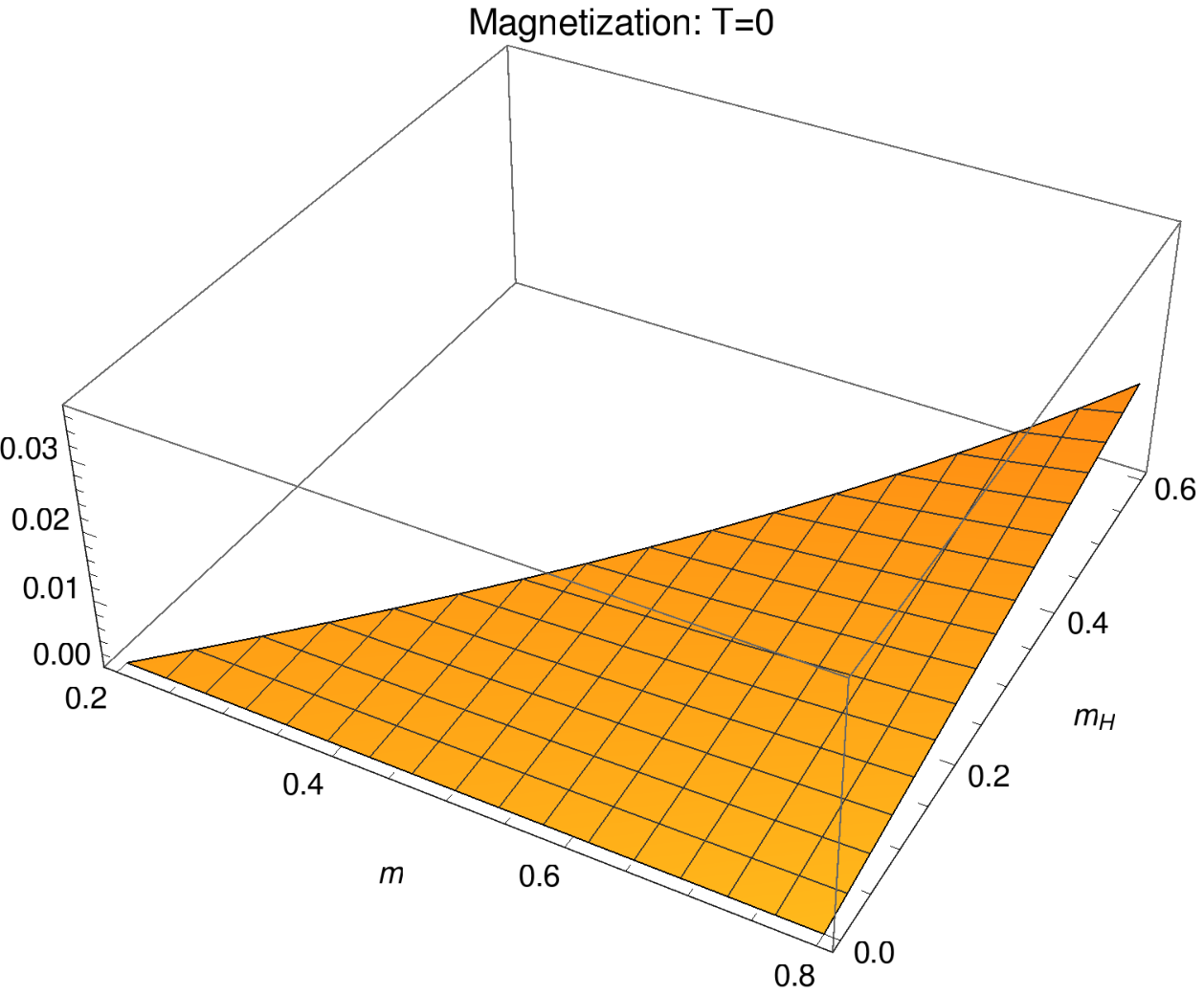} 
\includegraphics[width=8.0cm]{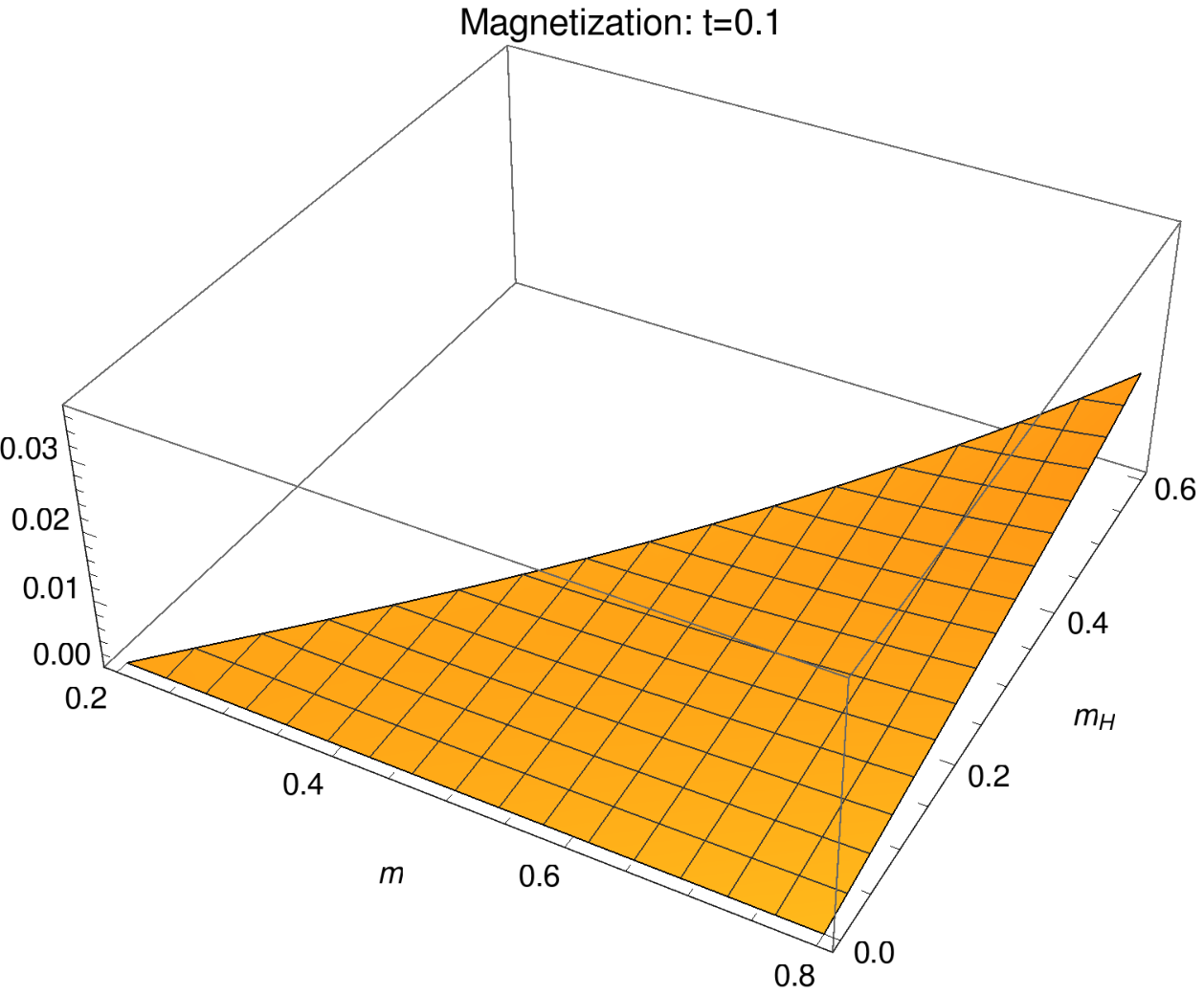}}
\vspace{7mm}
\hbox{
\includegraphics[width=8.0cm]{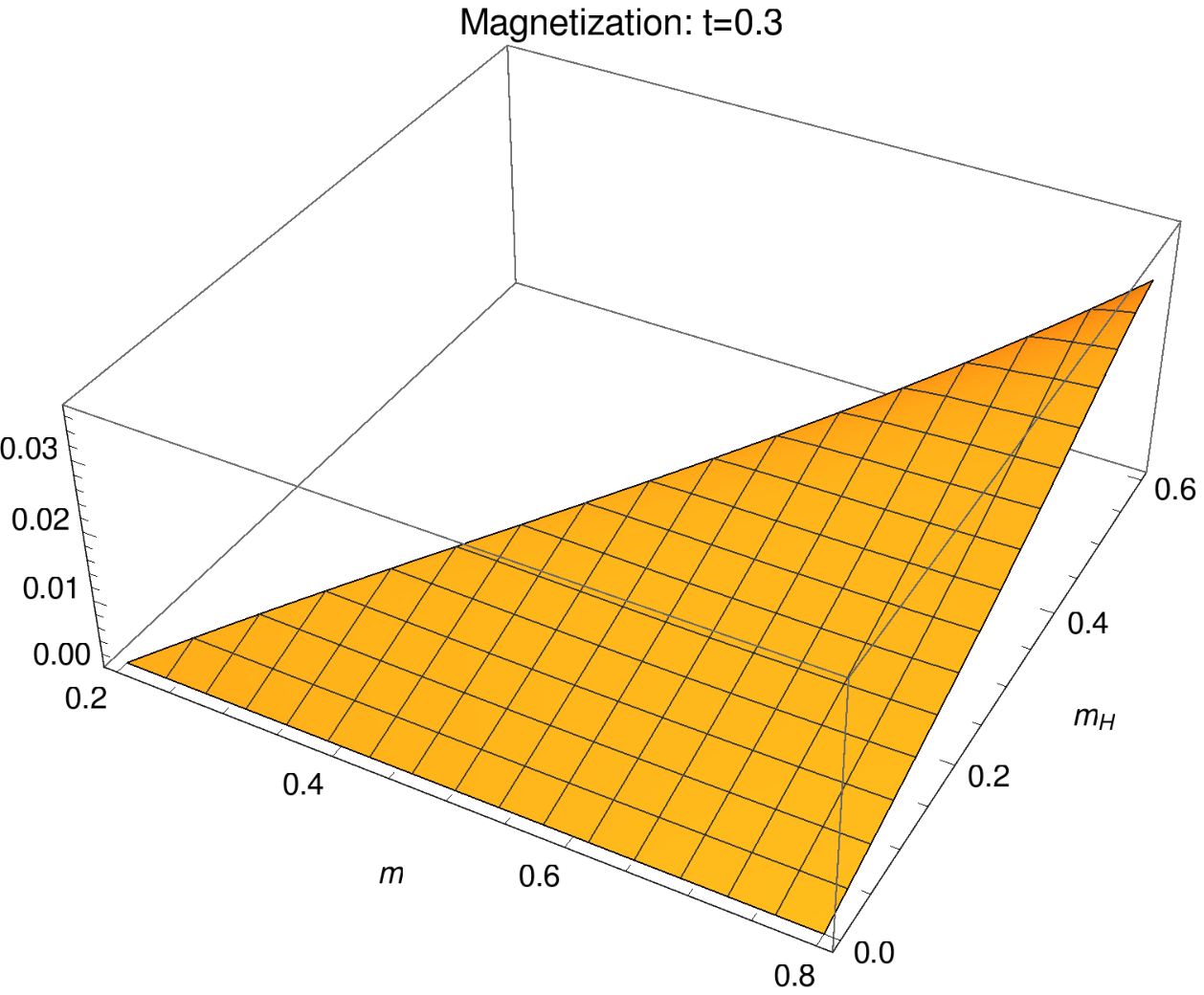}
\includegraphics[width=8.0cm]{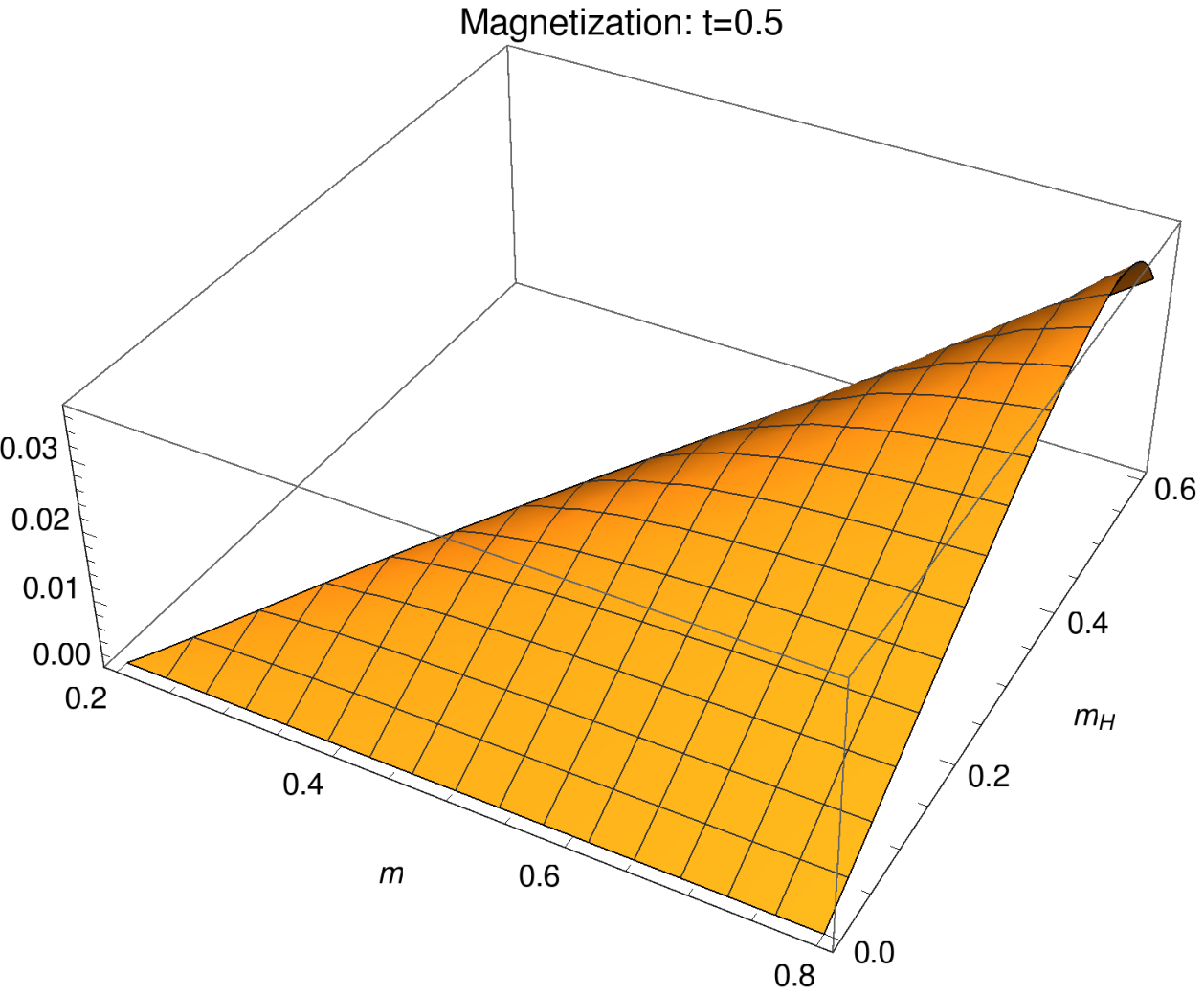}}
\end{center}
\caption{[Color online] Magnetization $M(T,H_s,H)$ at zero and finite temperature as a function of magnetic ($m_H$) and staggered ($m$) field
strength for the square-lattice antiferromagnet. The upper left figure refers to $T$=0, the other figures refer to the temperatures
$t = T/2 \pi \rho_s = \{ 0.1, 0.3, 0.5 \}$ (left to right, top to bottom).}
\label{figure4}
\end{figure}

A plot for the zero-temperature case is shown on the upper left of Fig.~\ref{figure4}. We observe a positive magnetization $M(0,H_s,H)$
in the direction of the magnetic field, that increases when magnetic and staggered fields become stronger. Again, these effects can be
explained by suppression of quantum fluctuations. The staggered field alone cannot induce a magnetization because it suppresses
fluctuations of up-spins and down-spins in the same manner. By incorporating a magnetic field, however, an asymmetric situation is
generated: the net external field ($H_s+H$) pointing up on sublattice $A$, is stronger than the net external field ($H_s-H$) pointing down
on sublattice $B$. As a consequence, the magnetization takes positive values in the direction of the magnetic field, because quantum
fluctuations on the $A$-sublattice are more suppressed. In stronger magnetic and staggered fields the value of the magnetization, as the
plot indicates, still is rather small, approximately $M(0,H_s,H) \approx 0.03 / a^2$. This is because we are describing a two-loop effect.

Let us finally investigate the effects caused by finite temperature. Along with the $T$=0 contribution, in Fig.~\ref{figure4}, we depict
the total magnetization $M(T,H_s,H)$ for the three temperatures $T/2 \pi \rho_s = \{ 0.1, 0.3, 0.5 \}$ (left to right, top to
bottom).\footnote{Notice that the units for the total magnetization $M(T,H_s,H)$ in the figures are the same as the units for the staggered
magnetization, namely $1/a^2$.} At very low temperatures, the magnetization is barely unchanged as compared to the $T$=0 magnetization.
However, one already notices the quite counterintuitive phenomenon that becomes apparent at more elevated temperatures: the total
magnetization grows when temperature is raised while keeping magnetic and staggered field strength fixed.

This comes quite unexpectedly because one would rather assume the total magnetization to drop as a consequence of the thermal fluctuations
that become stronger at higher temperatures. First of all we point out that the analogous phenomenon has also been observed in
three-dimensional antiferromagnets, subjected to magnetic fields aligned with the order parameter: according to Eq.~(7.4.126) of
Ref.~\citep{Nol86}, the magnetization grows when temperature increases. Moreover, in Ref.~\citep{AUW77} dealing with the thermal and
magnetic properties of a quasi two-dimensional antiferromagnet, the phenomenon has also been observed experimentally.

\begin{figure}
\begin{center}
\hbox{
\includegraphics[width=7.6cm]{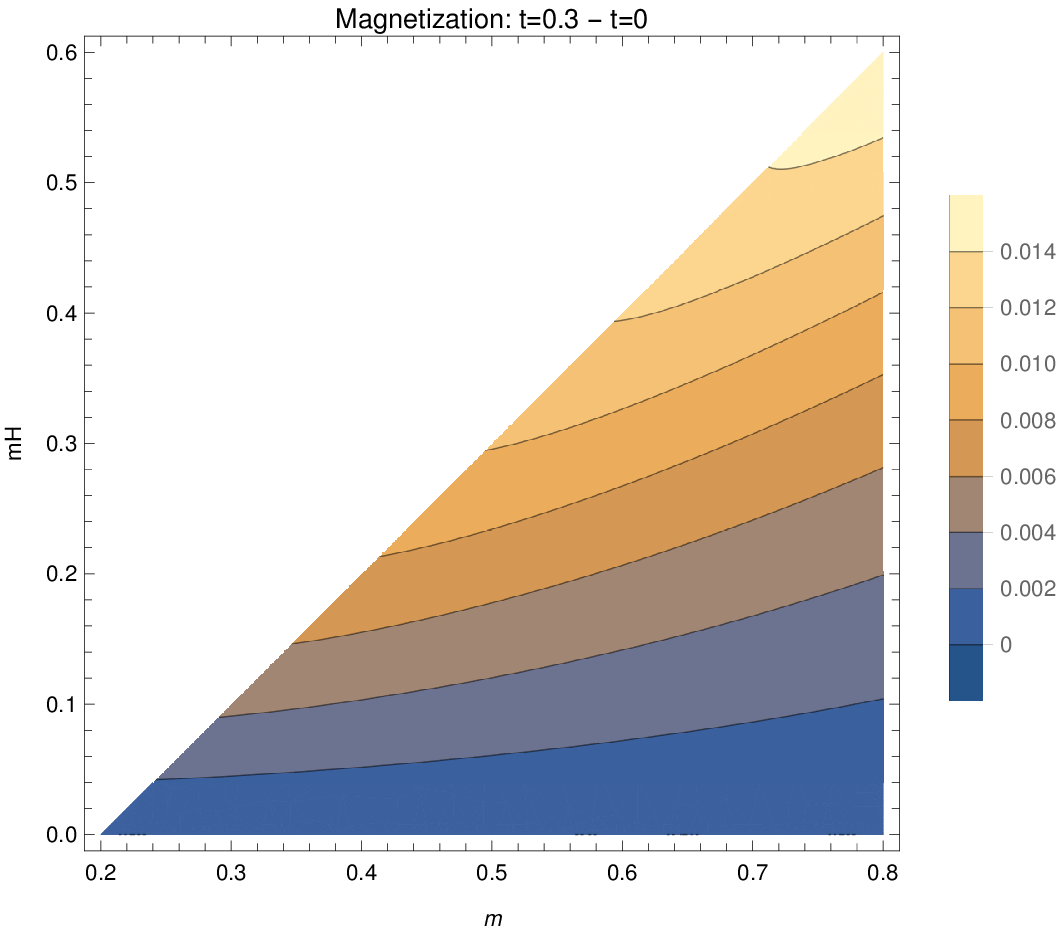} 
\includegraphics[width=7.6cm]{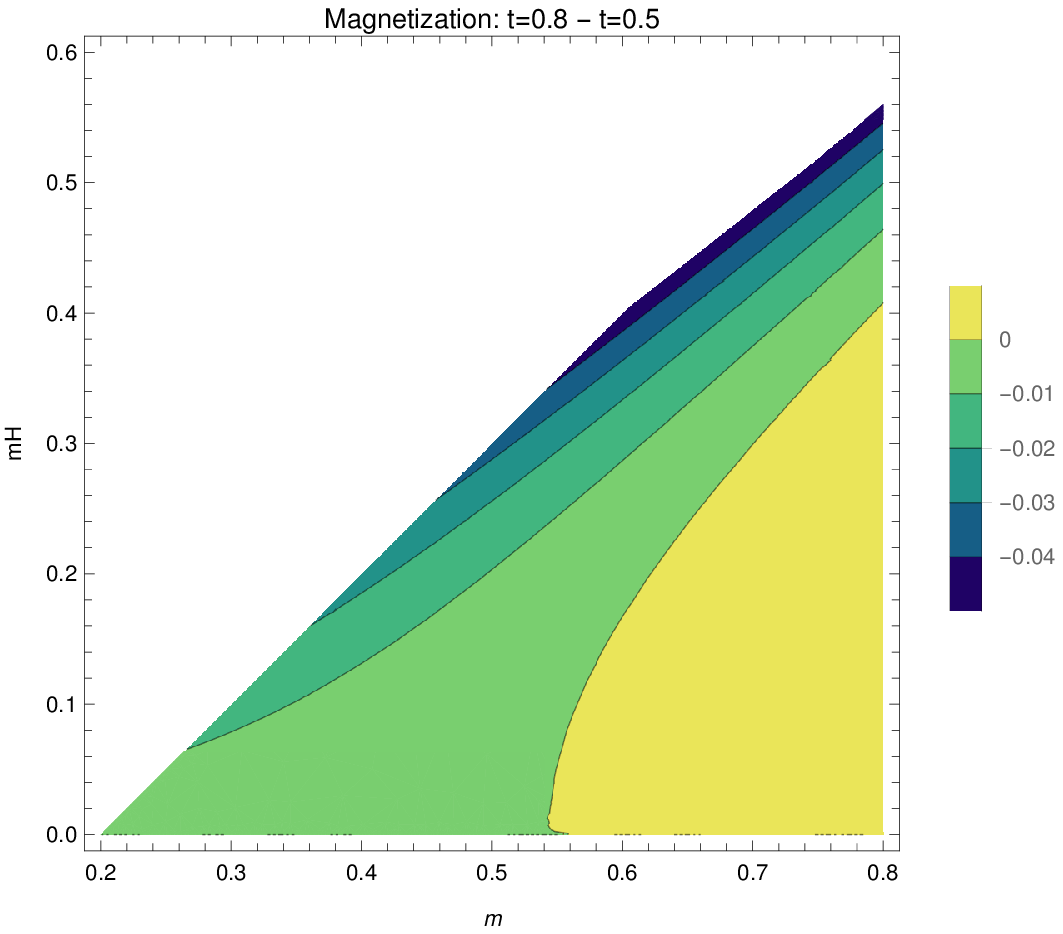}}
\end{center}
\caption{[Color online] Change of magnetization $M(T,H_s,H)$ going from $T$=0 to $t$=0.3 (left) and going from $t$=0.5 to $t$=0.8 (right),
as a function of magnetic ($m_H$) and staggered ($m$) field strength for the square-lattice antiferromagnet.}
\label{figure5}
\end{figure}

It should be emphasized that the magnetization does not monotonously increase with temperature. At more elevated temperatures, the
magnetization starts to drop. Indeed, this is the response one would intuitively expect. This is illustrated in Fig.~\ref{figure5} where we
provide contour plots for the two cases (a) magnetization going from zero temperature to $t$=0.3, and (b) going from $t$=0.5 to $t$=0.8. As
witnessed by case (a), at low temperatures the magnetization increases (with respect to $T$=0) in the entire parameter region that we
consider. However, at more elevated temperatures, going from $t$=0.5 to $t$=0.8 as in scenario (b), the magnetization starts to drop
because thermal fluctuations become stronger.

\section{Conclusions}
\label{conclusions}

The low-energy behavior of antiferromagnetic films subjected to a magnetic field aligned with the order parameter has been analyzed
systematically within the framework of magnon effective field theory. Low-temperature representations for the free energy density, the
staggered magnetization, and the magnetization have been derived up to two-loop order. In our numerical analysis we focused on the
square-lattice antiferromagnet where all relevant low-energy effective couplings are known from Monte Carlo simulations.

Considering the free energy density, we have illustrated that the two-loop correction is small with respect to the dominant one-loop
free Bose gas contribution. At zero temperature, the order parameter increases when the magnetic and staggered fields become stronger.
While the magnetization at $T$=0 follows a similar pattern, it should be noted that the staggered field alone cannot induce any
magnetization. These observations can be understood in terms of suppression of quantum fluctuations by the external fields. The enhancement
of the order parameter due to the magnetic field is reminiscent of magnetic catalysis.

At finite temperature, the staggered magnetization decreases due to thermal fluctuations -- as one would intuitively expect. What comes
quite as a surprise is that the total magnetization initially grows when temperature is raised while keeping magnetic and staggered field
strength fixed. At more elevated temperatures, however, total magnetization starts to decrease.

We emphasize that our effective field theory predictions for the square-lattice antiferromagnet are parameter-free -- both at zero and
finite temperature. At $T$=0, the relevant combination of next-to-leading order effective constants $k_2 + k_3$ is known from loop-cluster
Monte Carlo simulations. At finite temperature, such next-to-leading order effective constants only show up beyond two loops: the
thermodynamic properties of antiferromagnetic films on a bipartite lattice in general -- not restricted to the square lattice -- are fully
determined by the leading-order effective constants $\rho_s$ (spin stiffness) and $M_s$ (order parameter). In this sense, the rather
counterintuitive behavior of the magnetization exhibited by antiferromagnetic films in magnetic fields aligned with the order parameter,
is universal.

\section*{Acknowledgments}
The author thanks W.\ Nolting for correspondence.

\begin{appendix}

\section{Explicit Calculations}
\label{appendixA}

In this appendix we provide some additional material concerning the evaluation of the Feynman graphs for the free energy density.

\subsection{One-loop contribution to the free energy density}
\label{appendixA1}

Evaluating the one-loop graph in effective field theory (graph $3$ of Fig.~\ref{figure1}) boils down to evaluating the functional integral
$J$
\begin{equation}
J = \int [ \mbox{d} U ] \exp \! \Big[ - \int \! \mbox{d}^d x \, {\cal L}_{kin} \Big] \, , 
\end{equation}
which yields the one-loop free energy density $z_3$ via
\begin{equation}
z_3 = - \frac{1}{V_d} \, \log J \, , 
\end{equation}
where $V_d$ is the Euclidean volume.

To this end we consider the derivative of $J$ with respect to the magnon mass squared,
\begin{equation}
\frac{\partial}{\partial M^2} \, J = - \int [ \mbox{d} U ] \exp \! \Big[ - \int \! \mbox{d}^d x \, {\cal L}_{kin}
\Big] \, \frac{\rho_s}{2} \! \int \! \mbox{d}^d x \, U^a U^a \, ,
\end{equation}
where the kinetic term for the magnons in Euclidean space is
\begin{equation}
{\cal L}_{kin} = \mbox{$ \frac{1}{2}$} \rho_s \partial_{\mu} U^a \partial_{\mu} U^a + \mbox{$ \frac{1}{2}$} \rho_s M^2 U^a U^a
+ i \rho_s H \epsilon_{ab} \partial_0 U^a U^b - \mbox{$ \frac{1}{2}$} \rho_s H^2 U^a U^a \, .
\end{equation}
Using the physical magnon fields $u(x)$ and $u^{*}(x)$ defined in Eq.~(\ref{physicalMagnons}), we obtain
\begin{eqnarray}
\frac{\partial}{\partial M^2} \, J & = & - \int [ \mbox{d} u ] [ \mbox{d} u^{*} ] \exp \! \Big[ - \int \! \mbox{d}^d x
\, {\cal L}_{kin} \Big] \, \frac{\rho_s}{2} \! \int \! \mbox{d}^d x \, u u^{*} \nonumber \\
& & = - \frac{V_d J}{2} \, \Big\{ G^+(0) + G^-(0) \Big\} \nonumber \\
& & = - V_d J {\hat G}(0) \, ,
\end{eqnarray}
with
\begin{equation}
{\cal L}_{kin} = \mbox{$ \frac{1}{2}$} \rho_s \partial_{\mu} u \partial_{\mu} u^{*} + \mbox{$ \frac{1}{2}$} \rho_s M^2 u u^{*}
- \mbox{$ \frac{1}{2}$} \rho_s H (u^{*} \partial_0 u - u \partial_0 u^{*}) - \mbox{$ \frac{1}{2}$} \rho_s H^2 u u^{*} \, .
\end{equation}
In Euclidean space, the thermal propagator $G^{+}(x)$ -- referring to magnon $u(x)$ -- and the thermal propagator $G^{-}(x)$ -- referring to
magnon $u^{*}(x)$ -- are defined as
\begin{equation}
G^{\pm}(x) = \sum_{n = - \infty}^{\infty} \Delta^{\pm}({\vec x}, x_4 + n \beta) \, , \qquad \beta = \frac{1}{T} \, .
\end{equation}
At the origin $x$=0, the thermal propagators coincide,
\begin{equation}
G^{+}(0) = G^{-}(0) \equiv {\hat G}(0) \, .
\end{equation}
The explicit expression is provided in Eq.~(\ref{ThermalPropagatorsJacobi}).

Collecting partial results, the one-loop free energy density amounts to
\begin{equation}
z_3 = - {\hat g}_0 - \frac{M^{3/2}_s H^{3/2}_s}{6 \pi \rho_s^{3/2}} \, .
\end{equation}
Note that we have used Eqs.~(\ref{ThermalPropagatorsg1}) and (\ref{derg0}).

\subsection{Two-loop contribution to the free energy density}
\label{appendixA2}

Let us evaluate the two-loop graph $4b$ of Fig.~\ref{figure1}. The terms quartic in the magnons fields resulting from the leading-order
effective Lagrangian ${\cal L}^2_{eff}$ are
\begin{equation}
{\cal L}_{4b} = \mbox{$ \frac{1}{4}$} \rho_s \partial_{\mu} u \partial_{\mu} u^{*} u u^{*}
+ \mbox{$ \frac{1}{8}$} \rho_s \partial_{\mu} u^{*} u \partial_{\mu} u^{*} u
+ \mbox{$ \frac{1}{8}$} \rho_s \partial_{\mu} u u^{*} \partial_{\mu} u u^{*}
+ \mbox{$ \frac{1}{8}$} \rho_s M^2 u u^{*} u u^{*} \, .
\end{equation}
Evaluating the functional integral
\begin{equation}
J_{4b} = \int [ \mbox{d} u ]  [ \mbox{d} u^{*} ] \exp \! \Big[ - \int \! \mbox{d}^d x \, {\cal L}_{kin} \Big] \, \! \int \! \mbox{d}^d x \,
{\cal L}_{4b} \, ,
\end{equation}
the respective contribution to the free energy density is
\begin{eqnarray}
\label{z4bAppendix}
z_{4b} & = & \frac{H}{2 \rho_s} {\Big( {\dot G}^+(x) - {\dot G}^-(x) \Big)}_{|x=0} {\Big( G^+(x) + G^-(x) \Big)}_{|x=0}
+ \frac{H^2}{4 \rho_s} {\Big( G^+(x) + G^-(x)  \Big)}^2_{|x=0} \nonumber \\
& = & \frac{H}{\rho_s} \, {\hat g}_1 \, \frac{\partial {\hat g}_0}{\partial H}
- \frac{\sqrt{M_s H_s} H}{4 \pi \rho_s^{3/2}} \, \frac{\partial {\hat g}_0}{\partial H}
- \frac{H^2}{\rho_s}{( {\hat g}_1)}^2
+ \frac{\sqrt{M_s H_s} H^2}{2 \pi \rho_s^{3/2}} \, {\hat g}_1 - \frac{M_s H_s H^2}{16 \pi^2 \rho_s^2} \, .
\end{eqnarray}
In the course of the calculation we have used the fact that the thermal propagators obey the equations\footnote{We are in Euclidean
space with Euclidean time coordinate $x_4 = i t$.}
\begin{equation}
\Big\{ \Box - M^2 \pm 2 H \partial_{x_4} + H^2 \Big\} \, G^{\pm}(x)_{|_{x=0}} = 0 \, .
\end{equation}
Whereas single time derivatives of the thermal propagators at the origin $x=0$ vanish when no magnetic field is present,
\begin{equation}
{\dot G^{\pm}}(x)_{|_{x=0}} = 0 \, , \qquad \qquad \qquad \qquad (H = 0) \, ,
\end{equation}
this is different in nonzero magnetic fields. Starting with the representation for the thermal propagator, Eq.~(\ref{ThermalPropagators3}),
\begin{equation}
G^{\pm}(x) = \frac{1}{2 \sqrt{\pi}} \, \sum_{n = - \infty}^{\infty} \, {\int}_{\!\!\!0}^{\infty} \mbox{d} \lambda \, \int
\frac{{\mbox{d}}^{d_s} p}{(2 \pi)^{d_s}} \lambda^{-\frac{1}{2}} e^{-\lambda({\vec p \,}^2 + M^2)} e^{i {\vec p} \, {\vec x}} e^{-\frac{{(x_4 + n \beta)}^2}{4 \lambda}}
e^{\mp H(x_4 + n \beta)} \, ,
\end{equation}
we obtain the Euclidean time derivatives, evaluated at the origin $x=0$, as
\begin{eqnarray}
\frac{\partial}{\partial x_4} \, G^{\pm}(x)_{|x=0} & = & - \frac{\beta}{4 \sqrt{\pi}} \, \sum_{n = - \infty}^{\infty} \, {\int}_{\!\!\!0}^{\infty}
\mbox{d} \lambda \, \int \frac{{\mbox{d}}^{d_s} p}{(2 \pi)^{d_s}} \lambda^{-\frac{3}{2}} e^{-\lambda({\vec p \,}^2 + M^2)} \, n \,
e^{-\frac{n^2 \beta^2}{4 \lambda}} e^{\mp n \beta H} \nonumber \\
& & \hspace{-1.2truecm}\mp \frac{H}{2 \sqrt{\pi}} \, \sum_{n = - \infty}^{\infty} \, {\int}_{\!\!\!0}^{\infty} \mbox{d} \lambda \, \int
\frac{{\mbox{d}}^{d_s} p}{(2 \pi)^{d_s}} \lambda^{-\frac{1}{2}} e^{-\lambda({\vec p \,}^2 + M^2)} e^{-\frac{n^2 \beta^2}{4 \lambda}} e^{\mp n \beta H} \, .
\end{eqnarray}
The first contribution, using the identity,
\begin{equation}
e^{-\frac{n^2 \beta^2}{4 \lambda}} n \, e^{\mp n \beta H} = \mp \frac{1}{\beta} \, \frac{\partial}{\partial H} \Bigg\{ e^{-\frac{n^2 \beta^2}{4 \lambda}}
e^{\mp n \beta H}  \Bigg\} \, ,
\end{equation}
can be expressed in terms of the Jacobi theta function, defined by Eq.~(\ref{Jacobi3}). After a few trivial manipulations we end up with
\begin{equation}
\frac{\partial}{\partial x_4} \, G^{\pm}(x)_{|x=0} = \pm \frac{1}{2} \frac{\partial}{\partial H} g^{\pm}_0 \mp H g^{\pm}_1
\pm \frac{H M}{4 \pi} \, ,
\end{equation}
which finally leads to the result Eq.~(\ref{z4bAppendix}).

\end{appendix}

\end{document}